\documentclass[12pt]{article}

\usepackage{graphicx}
\usepackage{amssymb,amsmath, color}
\usepackage{amsthm}
\usepackage{soul}
\usepackage{comment}

\newtheorem{theorem}{Theorem}
\newtheorem{corollary}{Corollary}
\newtheorem{lemma}{Lemma}
\newtheorem{remark}{Remark}

\newcommand{\blue}[1]{\textcolor{blue}{#1}}

\begin{document}

\title{\bf Homogeneity of multinomial populations when data are classified into a large number of groups}

\author{
M.V. Alba-Fern\'andez$^{1}$\thanks{CONTACT M.V. Alba-Fern\'andez. Email: mvalba@ujaen.es},  M.D. Jim\'{e}nez--Gamero$^{2}$, F.J. Ariza-L\'opez$^{3}$ \\
\small{$^{1}$Dpt. Statistics and Operations Research, University of Ja\'en,  Spain} \\
\small{$^{2}$Dpt. Statistics and Operations Research, University of Seville,  Spain} \\
\small{$^{3}$Dpt. Engineering Cartography, Geodesy and Photogrammetry, University of Ja\'en,  Spain}}

\maketitle

\noindent{\bf Abstract.} Suppose that we are interested in the comparison of two independent categorical variables. Suppose also that the population is divided into subpopulations or groups. Notice that the distribution of the target variable may vary across subpopulations, moreover, it may happen that the two independent  variables have the same distribution in the whole population, but their distributions could differ in some groups. So, instead of testing the homogeneity of the two categorical variables, one may be interested in simultaneously testing the homogeneity in all groups.
A novel procedure is proposed for carrying out such a testing problem. The test statistic is shown to be asymptotically normal, avoiding the use of complicated resampling methods to get $p$-values. Here by asymptotic we mean when the number of groups increases; the sample sizes of the data from each group can either  stay bounded or grow with the number of groups. The finite sample performance of the proposal is empirically evaluated through an extensive simulation study. The usefulness of the proposal is illustrated by three data sets coming from  diverse experimental fields such as education, the COVID-19 pandemic and digital elevation models.


\section{Introduction}\label{Introduction}
\subsection{Motivation}\label{motivacion}
The two-sample problem for categorical data is a classical one in Statistics. 
Suppose that the population is divided into subpopulations or  groups. Notice that the distribution of the target variable may vary across subpopulations, moreover, it may happen that the two categorical  variables have the same distribution in the whole population, but their distributions could differ in some groups. So, instead of testing the homogeneity of the two categorical variables, one may be interested in simultaneously testing the homogeneity in all groups. For example, assume that we would like to study whether men and women have the same opinion on a certain topic, based on a qualitative response.
 In this setting, the age of respondents may be a key factor in explaining latent differences in their opinions. So, it seems reasonable to develop a formal procedure to test whether, for all age groups, the percentages of categorised opinions are the same for men and women or not.

Motivated by the above example and others with a similar structure, this piece of work tackles the problem of testing the equality of two categorical independent variables when there is a moderate to large number of (disjoint) subsets of the target population, and the user is interested in simultaneously testing the equality of the two categorical variables in each group.
Assume that independent samples are drawn from each population at each subgroup. The reader can recognize that the problem is equivalent to that of testing a  number, say $k$, of multinomial two-sample problems. $k$ represents the number of subgroups (age groups in the motivating example). Specifically, assume we have
$k$ independent pairs of multinomial data, all of them with the same number of categories or classes, $d \, (\geq 2)$,
\begin{eqnarray*}
{\boldsymbol N}_{1r}=(n_{1r1}, \ldots, n_{1rd})^\top & \sim &  \mathcal{M}(n_{1r.}; {\boldsymbol \pi}_{1r}),\\
{\boldsymbol N}_{2r}=(n_{2r1}, \ldots, n_{2rd})^\top & \sim &  \mathcal{M}(n_{2r.}; {\boldsymbol \pi}_{2r}),\quad 1 \leq r \leq k,
\end{eqnarray*}
with ${\boldsymbol \pi}_{ir}=(\pi_{ir1}, \ldots, \pi_{ird})^\top\in \Xi_{d}=\{(\pi_1,\ldots, \pi_d)^\top:\; \pi_j  \geq 0, \; 1\leq j \leq d, \; \sum_{j=1}^d \pi_j=1\}$, $i=1,2$, $1 \leq r \leq k$. ${\boldsymbol N}_{1r}$ and ${\boldsymbol N}_{2r}$ are also assumed to be independent for each $1 \leq r \leq k$.

We are interested in testing
\begin{eqnarray*}
H_0:  & {\boldsymbol \pi}_{1r}={\boldsymbol \pi}_{2r},& 1 \leq r \leq k, \\
H_1:  & {\boldsymbol \pi}_{1r} \neq {\boldsymbol \pi}_{2r},& \mbox{ for some }r,
\end{eqnarray*}
for large $k$. In terms of the motivating example, ${\boldsymbol N}_{1r}$ and ${\boldsymbol N}_{2r}$ are the vectors of observed frequencies of the categories of the key variable in the sample of subgroup (age) $r$ taken from men --population 1-- and women --population 2--, respectively;  such samples have sizes $n_{1r.}$ and $n_{2r.}$, respectively;  ${\boldsymbol \pi}_{1r}$  and ${\boldsymbol \pi}_{2r}$ are  the vectors of unknown true proportions of the categories of the key variable in subgroup (age) $r$ of men and women, respectively, where $1 \leq r \leq k$.

Two classical approaches for getting a test of $H_0$ vs $H_1$ are the likelihood ratio test (LRT) and the union-intersection test (UIT). We will see that when the sample sizes, $n_{11.}, n_{21.}, \ldots n_{1k.}, n_{2k.}$,   remain bounded or increase with $k$,
 but at a lower rate than $k$, those methodologies  do not yield feasible tests. Therefore, a new procedure is needed for testing $H_0$ vs $H_1$.

\subsection{Related work}

Let $X_{11},X_{21}, \ldots, X_{1k},X_{2k}$ be random variables so that $P(X_{ir}=j)=\pi_{irj}$, $1\leq j \leq d$, $1 \leq r \leq k$, $i=1,2$, and let
${\boldsymbol \zeta}_{11}, {\boldsymbol \zeta}_{21} \ldots, {\boldsymbol \zeta}_{1k},{\boldsymbol \zeta}_{2k}$ be random vectors defined as
$$ % \begin{equation}\label{zeta}
 {\boldsymbol \zeta}_{ir}=(I(X_{ir}=1), \ldots, I(X_{ir}=d))^\top \in \{0,1\}^d,
$$ % \end{equation}
 $1 \leq r \leq k$, $i=1,2$.  With this notation, ${\boldsymbol N}_{ir}=\sum_{s=1}^{n_{ir.}} {\boldsymbol \zeta}_{irs}$, where ${\boldsymbol \zeta}_{ir1}, \ldots, {\boldsymbol \zeta}_{irn_{ir.}}$ are independent and identically distributed (iid) from   ${\boldsymbol \zeta}_{ir}$,  $1 \leq r \leq k$, $i=1,2$, and
    the null hypothesis  can be rewritten as  $H_0:\, E({\boldsymbol \zeta}_{1r})=E({\boldsymbol \zeta}_{2r})$, $ 1 \leq r \leq k$, or equivalently,
 $$H_0:\, E({\boldsymbol \zeta}_{1})=E({\boldsymbol \zeta}_{2}),$$
where ${\boldsymbol \zeta}_{i}=({\boldsymbol \zeta}_{i1}^\top, \ldots, {\boldsymbol \zeta}_{ik}^\top)^\top \in \{0,1\}^{dk}$, $i=1,2$, that is, $H_0$ can be seen as the equality of the means of two high-dimensional vectors. This problem has been studied in the statistical literature and several tests have been proposed, under certain assumptions. Examples are the tests in \cite{Bai&Saranadasa,Chen&Qin, Park&Ayyala2013, Srivastavaetal}.  The procedures in \cite{Bai&Saranadasa,Chen&Qin,Park&Ayyala2013} assume that the data satisfy a factor model that is not met in our setting. The test in \cite{Srivastavaetal} assumes that the sample sizes increase with $k$ at a certain rate. As will be shown later, our procedure is valid even if the sample sizes stay bounded.
\cite{Plunkett&Park} proposed a test for testing the equality of the mean of two high-dimensional multinomial populations. In our setting the vectors ${\boldsymbol \zeta}_{1}$ and ${\boldsymbol \zeta}_{2}$ are compounds of independent multinomial subvectors, but the entire vectors have not multinomial distributions. 
Apparently, our objective has not been dealt with in the literature so far.

\subsection{Outline}

Section \ref{LRT&UIT}  calculates the  UIT and the LRT of $H_0$ vs $H_1$. Through simulations, it is shown that feasible versions of those tests fail to keep the level.
A novel procedure for simultaneously testing $k$ multinomial two-sample problems is proposed. Section \ref{The.test.statistic} is devoted to the construction of the test statistic.  For each $1 \leq r \leq k$, an unbiased estimator of the  Euclidean distance between the probability vectors
 ${\boldsymbol \pi}_{1r}$ and ${\boldsymbol \pi}_{2r}$ is derived.  Then, these estimators are combined to get the test statistic.  Because the null distribution of the proposed test statistic is unknown, we calculate its asymptotic distribution, as an approximation to its actual distribution. The asymptotic distribution turns out to be normal, with variance depending on the unknown values of the probability vectors. Here by asymptotic, we mean when the number of groups increases; the sample sizes of the data from each group can either stay bounded or grow with the number of groups. Three variance estimators are studied, which give rise to three tests of $H_0$. All theoretical derivations, such as those required to get the asymptotic distribution of the test statistic and the study of the asymptotic power of the tests that results when considering each variance estimator,  are relegated to the Appendix.
 
The rest of the paper is organized as follows: Section \ref{simulation} summarizes the results of a large simulation experiment carried out to evaluate the finite sample performance of the tests, concerning both the level and the power; and  Section \ref{application} details three applications to data sets in different areas:  education level in men and women,   level of information of Spanish people about the COVID-19 vaccine, and digital elevation models over the zone of Allo (Navarra, Spain). Finally, Section
 \ref{conclusions} concludes.

Before ending this section we introduce some notation:
 all limits in this paper are taken when   $k \rightarrow \infty$;
 $\stackrel{\mathcal{L}}{\rightarrow}$ denotes convergence in
distribution; $\stackrel{P}{\rightarrow}$ denotes convergence in
probability; all vectors are column vectors; the superscript $^\top$ denotes
transpose; if $\boldsymbol x \in \mathbb{R}^d$, with $\boldsymbol x^\top=(x_1, \ldots,
x_d)$, then  ${\rm diag}(\boldsymbol x)$ is the $d \times d$ diagonal matrix whose
$(i,i)$ entry is $x_i$, $1\leq i \leq d$, $\Sigma_{ \boldsymbol x}={\rm diag}(\boldsymbol x)-\boldsymbol x \boldsymbol x^\top$, and $\| \boldsymbol x\|$ denotes the Euclidean norm of $\boldsymbol x$; ${\rm tr}$ stands for the trace (of a matrix); $E_0$ and $var_0$ denote expectation and variance, respectively, under $H_0$.

\section{The union-intersection test and the likelihood ratio test} \label{LRT&UIT}

The UIT  rejects $H_0$ for large values of the statistic
 ${\cal T}=\sum_{r=1}^k T_r$ (see Subsection \ref{UIT}, in the Appendix), where 
$T_{r}$ denotes the classical chi-square test statistic for testing $H_{0r}:  \, {\boldsymbol \pi}_{1r}={\boldsymbol \pi}_{2r}$ vs
$H_{1r}:  \, {\boldsymbol \pi}_{1r} \neq {\boldsymbol \pi}_{2r}$ (see, e.g. Pardo \cite{pardo2006libro} for  that test and some extensions). $T_r$   is asymptotically distributed under $H_{0r}$ as a $\chi^2_{d-1}$, a chi-square distribution with $d-1$ degrees of freedom, where here asymptotically means that $n_{1r.}$ and $n_{2r.}$  
%increase arbitrarily
are arbitrarily large, $1\leq r \leq k.$ Thus,  for large $k$ and under $H_0$, the test statistic
\begin{equation} \label{test.chi}
W_k=\frac{1}{\sqrt{k}}\frac{\sum_{r=1}^k(T_{r}-d+1)}{\sqrt{2(d-1)}}
\end{equation}
is approximately distributed as a standard normal distribution. A key assumption to derive such approximation is that $\displaystyle \min_{\, 1 \leq r \leq k , \, i=1,2} n_{ir.}  \to \infty$. 
%n_{1r.} \wedge   n_{2r.} \to \infty$.
On the contrary, in this paper, the interest focuses on testing $H_0$ when the sample sizes ($n_{11.}, n_{21.}, \ldots, n_{1k.}, n_{2k.} $) either remain bounded or increase with $k$,
 but at a lower rate than $k$, which seems to be a more realistic assumption when $k$ is large.
 %(although the proposed procedures are also valid for sample sizes increasing at the same rate as or even larger than $k$).
 In such a setting, the normal approximation to the null distribution of $W_k$ does not work. To numerically appreciate this fact, we carried out the following simulation experiment:  $k$ pairs of random samples were generated from populations  $\mathcal{M}(n_{1r.}; {\boldsymbol \pi}_{1r})$ and $\mathcal{M}(n_{2r.}; {\boldsymbol \pi}_{2r})$, with ${\boldsymbol \pi}_{1r}={\boldsymbol \pi}_{2r}=(1/d, \ldots, 1/d)$, $1 \leq r \leq k$, for $k=20,50,100,200,500,750$,  $(n_{1r.},n_{2r.})=(5,10)$, (10,10), (20,30), (30,30),  and $d=5,10,20$.
In each case, we applied the test that reject $H_0$  when $W_k>z_{1-\alpha}$, for   $\alpha=0.05$, where %$W_k$ is as defined in \eqref{test.chi},
$\Phi(z_{1-\alpha})=1-\alpha$ and $\Phi$ stands for the cumulative distribution function of the standard normal distribution.
The experiment was repeated 10,000 times.  Table \ref{tab:8} reports the fraction of times that the null hypothesis was rejected, which estimates the probability of type I error. Looking at Table \ref{tab:8} we see that,
 in most cases, the estimated probability of type I error is far apart from the nominal value $0.05$. Close results are obtained if  $H_0$ is rejected when ${\cal T} \geq \chi^2_{k(d-1), 1-\alpha}$, where $\chi^2_{k(d-1), 1-\alpha}$ denotes the $(1-\alpha)$ percentile of the law 
 $\chi^2_{k(d-1)}$.

\begin{table}[h]
\centering
\begin{tabular}{ccccc|ccccc}
%\hline
      & &\multicolumn{3}{c}{$d$} & & & \multicolumn{3}{c}{$d$} \\    \hline
  $k$ & $n_{1r},n_{2r}$ & 5 & 10 & 20 & $k$ & $n_{1r},n_{2r}$ & 5 & 10 & 20 \\ \hline
  20  & 5,10  & 0.038 & 0     & 0     & 200 & 5,10  & 0.058 & 0     & 0     \\
      & 10,10 & 0.061 & 0     & 0     &     & 10,10 & 0.155 & 0     & 0     \\
      & 20,30 & 0.064 & 0.052 & 0     &     & 20,30 & 0.106 & 0.080 & 0     \\
      & 30,30 & 0.065 & 0.052 & 0.003 &     & 30,30 & 0.098 & 0.093 & 0     \\
  50  & 5,10  & 0.034 & 0     & 0     & 500 & 5,10  & 0.100 & 0     & 0     \\
      & 10,10 & 0.081 & 0     & 0     &     & 10,10 & 0.302 & 0     & 0     \\
      & 20,30 & 0.074 & 0.054 & 0     &     & 20,30 & 0.154 & 0.140 & 0     \\
      & 30,30 & 0.073 & 0.058 & 0.001 &     & 30,30 & 0.128 & 0.152 & 0     \\
  100 & 5,10  & 0.048 & 0     & 0     & 750 & 5,10  & 0.133 & 0     & 0     \\
      & 10,10 & 0.105 & 0     & 0     &     & 10,10 & 0.411 & 0     & 0    \\
      & 20,30 & 0.084 & 0.068 & 0     &     & 20,30 & 0.192 & 0.178 & 0    \\
      & 30,30 & 0.075 & 0.074 & 0     &     & 30,30 & 0.156 & 0.199 & 0    \\
  \hline
\end{tabular}\caption{Estimated type I error for the test with critical region $W_k>z_{0.95}$, with $W_k$  as in \eqref{test.chi}. } \label{tab:8}
\end{table}

The reason for the bad behavior of the normal approximation to the null distribution of $W_k$ is that we are taking  $E_0(T_r)=d-1$ and $var_0(T_r)=2(d-1)$, but those equalities hold  for very large sample sizes. Indeed, if instead of $W_k$, we consider
\begin{equation}\label{test.chi.unfeasible}
W_k'=
%\frac{\sum_{r=1}^k(T_{r}-E_0(T_{r}))}{ \sqrt{\sum_{r=1}^kvar_0(T_{r})}},
\sum_{r=1}^k(T_{r}-E_0(T_{r}))/ \left(\sum_{r=1}^kvar_0(T_{r}) \right)^{1/2},
\end{equation}
then the normal approximation works really well. To see this, we repeated the previous experiment for the test that reject $H_0$ when 
$W_k'>z_{0.95}$. Table \ref{tab:88} displays the results obtained. Looking at this table we observe that in all cases the empirical level is reasonably close to the nominal value 0.05. The main  point is that the test based on  $W_k'$ is unfeasible, as to calculate $E_0(T_{r})$ and $var_0(T_{r})$, $1 \leq r \leq k$, for small to moderate sample sizes, we need to know ${\boldsymbol \pi}_{1r}$ and ${\boldsymbol \pi}_{2r}$,  $1 \leq r \leq k$.
\begin{table}[h]
\centering

\begin{tabular}{ccccc|ccccc}
%\hline
      & &\multicolumn{3}{c}{$d$} & & & \multicolumn{3}{c}{$d$} \\    \hline
  $k$ & $n_{1r},n_{2r}$ & 5 & 10 & 20 & $k$ & $n_{1r},n_{2r}$ & 5 & 10 & 20 \\ \hline
  20  & 5,10  & 0.055 & 0.053 & 0.052 & 200  & 5,10  & 0.051 & 0.052 & 0.050    \\
      & 10,10 & 0.054 & 0.053 & 0.048 &      & 10,10 & 0.052 & 0.051 & 0.048     \\
      & 20,30 & 0.054 & 0.054 & 0.049 &      & 20,30 & 0.047 & 0.049 & 0.047     \\
      & 30,30 & 0.058 & 0.053 & 0.051 &      & 30,30 & 0.051 & 0.051 & 0.052     \\
  50  & 5,10  & 0.054 & 0.051 & 0.051 & 500  & 5,10  & 0.051 & 0.050 & 0.048  \\
      & 10,10 & 0.053 & 0.050 & 0.049  &     & 10,10 & 0.051 & 0.051 & 0.049     \\
      & 20,30 & 0.054 & 0.052 & 0.048  &      & 20,30 & 0.050 & 0.054 & 0.049     \\
      & 30,30 & 0.052 & 0.052 & 0.051 &      & 30,30 & 0.051 & 0.051 & 0.050     \\
  100 & 5,10  & 0.050 & 0.051 & 0.051 & 750  & 5,10  & 0.053 & 0.053 & 0.050 \\
      & 10,10 & 0.053 & 0.053 & 0.049 &      & 10,10 & 0.051 & 0.052 & 0.047   \\
      & 20,30 & 0.048 & 0.052 & 0.048 &      & 20,30 & 0.047 & 0.053 & 0.049   \\
      & 30,30 & 0.052 & 0.051 & 0.052 &      & 30,30 & 0.051 & 0.051 & 0.052   \\
  \hline
\end{tabular}
\caption{Estimated type I error for the test with critical region $W_k'>z_{0.95}$, with $W'_k$  as in \eqref{test.chi.unfeasible}. } 
\label{tab:88}

\end{table}

Routine calculations, show that the LRT rejects $H_0$ for large values  of ${\cal S}=-2\sum_{r=1}^k \log (\lambda_r)$, where $\lambda_r=\prod_{j=1}^d \hat{\pi}_{rj}^{n_{1rj}+n_{2rj}}/(\hat{\pi}_{1rj}^{n_{1rj}} \hat{\pi}_{2rj}^{n_{2rj}})$, $\hat{\pi}_{irj}=n_{irj}/n_{ir.}$, $\hat{\pi}_{rj}=(n_{1rj}+n_{2rj})/(n_{1r.}+n_{2r.})$, $i=1,2$, $1\leq j \leq d$, $1 \leq r \leq k$. $-2\log(\lambda_r)$
is asymptotically distributed under $H_{0r}$ as a $\chi^2_{d-1}$, 
where here asymptotically means for large $n_{1r.}$ and $n_{2r.}$. Thus, proceeding as with the UIT, one could consider a statistic similar to $W_k$ ($W_k'$) with $T_r$ replaced with $-2\log(\lambda_r)$,  $1\leq r \leq k$, say $V_k$ ($V_k'$). Looking at Tables \ref{tab:TRV1}  and \ref{tab:TRV2}  (located in Subsection \ref{level:LRT}, in the Appendix) we see that the same problem found for the UIT arises here too.    

Summarizing, the UIT and the LRT cannot be applied for testing $H_0$ vs $H_1$, for small to moderate sample sizes.

\section{The new test} \label{The.test.statistic}

As seen in Section  \ref{LRT&UIT}, a main drawback when dealing with the UIT is that the expected value under $H_{0r}$ of $T_r$ is an unknown quantity, $1\leq r \leq k$. This section proposes a new test of $H_0$ vs $H_1$  which avoids the above issue. With this aim, instead of $W_k$, we consider another test statistic, which is similar in spirit but, in contrast to $W_k$, it has a known expectation under $H_{0}$.

To start with, we revisit the problem of testing the equality of two probability vectors, that is, the case of $k=1$. For a moment and just to simplify notation, when $k=1$, the index $r$ will be skipped. So, we write
${\boldsymbol N}_{1}=(n_{11}, \ldots, n_{1d})^\top  \sim   \mathcal{M}(n_{1.}; {\boldsymbol \pi}_{1})$,
${\boldsymbol N}_{2}=(n_{21}, \ldots, n_{2d})^\top  \sim   \mathcal{M}(n_{2.}; {\boldsymbol \pi}_{2})$,
${\boldsymbol \pi}_{1}=({\pi}_{11}, \ldots, {\pi}_{1d})^\top$ and ${\boldsymbol \pi}_{2}=({\pi}_{21}, \ldots, {\pi}_{2d})^\top$. Let $T_{1}$ denote the classical chi-square test statistic for testing $H_{01}:\,   {\boldsymbol \pi}_{1}={\boldsymbol \pi}_{2}$, which is equivalent to $H_{01}:\,   \|{\boldsymbol \pi}_{1}-{\boldsymbol \pi}_{2}\|=0$.   $T_1$ can be expressed as follows
\begin{equation}\label{T1}
T_1=\frac{n_{1.} \times n_{2.}}{n}(\hat{{\boldsymbol \pi}}_1-\hat{{\boldsymbol \pi}}_2)^\top {\rm diag}^{-1}(\hat{{\boldsymbol \pi}}) (\hat{{\boldsymbol \pi}}_1-\hat{{\boldsymbol \pi}}_2),
\end{equation}
where $n=n_{1.}+n_{2.}$, $\hat{{\boldsymbol \pi}}_1={\boldsymbol N}_1/n_{1.}$, $\hat{{\boldsymbol \pi}}_2={\boldsymbol N}_2/n_{2.}$, and $\hat{{\boldsymbol \pi}}=({\boldsymbol N}_1+{\boldsymbol N}_2)/n$. The matrix ${\rm diag}^{-1}(\hat{{\boldsymbol \pi}})$ is introduced in the expression of $T_1$ to get an asymptotically distribution free test statistic.
 Here by asymptotic it is understood that $\min\{n_{1.}, n_{2.}\} \to \infty$.  However, as in our setting the sample sizes may stay bounded,  the matrix ${\rm diag}^{-1}(\hat{{\boldsymbol \pi}})$  plays no role.
 If one instead takes as test statistic %($V$-statistic)
\[
T_{V_1}=(\hat{{\boldsymbol \pi}}_1-\hat{{\boldsymbol \pi}}_2)^\top (\hat{{\boldsymbol \pi}}_1-\hat{{\boldsymbol \pi}}_2)=\|\hat{{\boldsymbol \pi}}_1-\hat{{\boldsymbol \pi}}_2\|^2,
\]
then, its the expected value is
\[
E(T_{V_1})=\|{\boldsymbol \pi}_1-{\boldsymbol \pi}_2\|^2+\sum_{j=1}^d \left(\frac{\pi_{1j}(1-\pi_{1j})}{n_{1.}}+ \frac{\pi_{2j}(1-\pi_{2j})}{n_{2.}}
\right).
\]
We see that $T_{V_1}$ is a biased estimator of $\|{\boldsymbol \pi}_1-{\boldsymbol \pi}_2\|^2$, even under the null hypothesis $H_{01}$, and that its bias is not negligible for small or moderate sample sizes. This is because  $T_{V_1}$ is a V-statistic.
In order to remove the bias, it will be calculated the associated $U$-statistic which, after some algebra (see Subsection \ref{u_est} of the Appendix), becomes
\begin{equation} \label{TU1}
T_{U_1}=\frac{n_{1.}}{n_{1.}-1}\hat{{\boldsymbol \pi}}_1^\top \hat{{\boldsymbol \pi}}_1-\frac{1}{n_{1.}-1}+ \frac{n_{2.}}{n_{2.}-1}\hat{{\boldsymbol \pi}}_2^\top \hat{{\boldsymbol \pi}}_2-\frac{1}{n_{2.}-1}-2 \hat{{\boldsymbol \pi}}_1^\top \hat{{\boldsymbol \pi}}_2 \, \,.
\end{equation}

Let us come back to the problem of testing $H_0$. Let $T_{U_r}$ denote the statistic $T_{U_1}$ calculated from ${\boldsymbol N}_{1r}$ and ${\boldsymbol N}_{2r}$, $1 \leq r \leq k$, and consider the following test statistic
\[
T_{U}=\frac{1}{\sqrt{k}}\sum_{r=1}^k T_{U_r}.
\]
whose expected value and variance are
\begin{eqnarray*}
E(T_{U}) & = & \frac{1}{\sqrt{k}} \sum_{r=1}^k \|{\boldsymbol \pi}_{1r}-{\boldsymbol \pi}_{2r}\|^2  \geq 0,\\ %:= \theta_k \geq 0,
var(T_{U}) & = & \frac{1}{k}\sum_{r=1}^k var( T_{Ur}),\\
var( T_{Ur}) & =  & \frac{4}{n_{1r.}} ({\boldsymbol \pi}_{1r}-{\boldsymbol \pi}_{2r})^\top\Sigma_{{\boldsymbol \pi}_{1r}} ({\boldsymbol \pi}_{1r}-{\boldsymbol \pi}_{2r})+
 \frac{4}{n_{2r.}} ({\boldsymbol \pi}_{1r}-{\boldsymbol \pi}_{2r})^\top\Sigma_{{\boldsymbol \pi}_{2r}} ({\boldsymbol \pi}_{1r}-{\boldsymbol \pi}_{2r})\\
  & & +\frac{2}{n_{1r.}(n_{1r.}-1)} {\rm tr}(\Sigma_{{\boldsymbol \pi}_{1r}}^2) + \frac{2}{n_{2r.}(n_{2r.}-1)} {\rm tr}(\Sigma_{{\boldsymbol \pi}_{2r}}^2) +
  \frac{4}{n_{1r.}n_{2r.}} {\rm tr}(\Sigma_{{\boldsymbol \pi}_{1r}} \Sigma_{{\boldsymbol \pi}_{2r}}).
 \end{eqnarray*}
Under $H_0$, the expression of the variance of $T_{U}$ will be denoted as
$var_0(T_{U})=\frac{1}{k}\sum_{r=1}^k var_0( T_{U_r})$, which  greatly simplifies, since all terms involving ${\boldsymbol \pi}_{1r}-{\boldsymbol \pi}_{2r}$  vanish.

Observe that $E(T_{U})=0$ if and only if $H_0$ is true. Since $E(T_{U})=0$ under  $H_0$ and  $E(T_{U})> 0$ under alternatives, the null hypothesis is rejected for large values of $T_{U}$. To determine what are ``large values'' we should know either its null distribution or at least a consistent approximation.
The null distribution of $T_{U}$ is clearly unknown. As an approximation, Theorem \ref{asymptotic.null.distribution}  in the Appendix shows that under certain mild assumptions, when
$H_0$ is true,
\begin{equation} \label{asin.normality}
T_{U}/\sqrt{var_0(T_{U})} \stackrel{\mathcal{L}}{\longrightarrow} Z,
\end{equation}
where $Z\sim N(0,1)$. Therefore, if $var_0(T_{U})$ were a known quantity, the test that rejects $H_0$ when
$T_{U}/\sqrt{var_0(T_{U})}  \geq z_{1-\alpha},$
for some $\alpha \in (0,1)$, would have (asymptotic) level $\alpha$.

Two conditions are required to get the asymptotic normality in \eqref{asin.normality}. One of them is commented in Remark
\ref{remark.trazas}, and the other is related to the sample sizes (see
  \eqref{sample.sizes}  in the Appendix), which states that the sample sizes must be comparable: either all of them small (bounded) or large (they are allowed to increase with $k$). This assumption is commonly adopted in the $k$-sample problem (see, e.g. \cite{matcom09} for categorical data, and \cite{ Alba2017, two, fast} for numerical data).

For the result in Theorem \ref{asymptotic.null.distribution}  to be useful to get a critical region for testing $H_0$, a consistent estimator of $var_0(T_{U})$ is needed. Next, three estimators of $var_0(T_{U})$  will be investigated, giving rise to three tests of $H_0$.

\subsection{ Test 1} \label{test.1}
$var_0(T_{U})$ is unknown because the quantities $\Sigma_{{\boldsymbol \pi}_{1r}}^2$,  $\Sigma_{{\boldsymbol \pi}_{1r}}^2$  and $\Sigma_{{\boldsymbol \pi}_{1r}} \, \Sigma_{{\boldsymbol \pi}_{2r}}$ are unknown.
Unbiased estimators of these quantities can be derived by using $U$-estimators as follows.

Assume for a moment that $r=1$ (this is unimportant). To simplify notation, we skip the subindex $r$. %As in the previous section,
Let
 $X_{11}, \ldots, X_{1n_{1.}}$ be iid so that $P(X_{11}=j)=\pi_{1j}$, $1\leq j \leq d$, and let $X_{21}, \ldots, X_{2n_{2.}}$ be iid and independent from the previous sample  so that $P(X_{21}=j)=\pi_{2j}$, $1\leq j \leq d$. Let ${\boldsymbol \pi}_1=( \pi_{11}, \ldots, \pi_{1d})^\top$, ${\boldsymbol \pi}_2=( \pi_{21}, \ldots, \pi_{2d})^\top$ and
$$
{\boldsymbol \zeta}_{ij}=(I(X_{ij}=1), \ldots, I(X_{ij}=d))^\top \in \{0,1\}^d,
 $$
 $1\leq j \leq n_{i.}$,  $i=1,2$. With this notation, an unbiased estimator of $ \Sigma_{{\boldsymbol \pi}_{i}}$ is
\begin{equation} \label{estim.Sigma}
\widehat{\Sigma}_{{\boldsymbol \pi}_{i}} =\frac{n_{i.}}{n_{i.}-1} \Sigma_{\hat{{\boldsymbol \pi}}_{i}}, \qquad \hat{{\boldsymbol \pi}}_{i}=\frac{1}{n_{i.}}{\boldsymbol N}_i, \qquad {\boldsymbol N}_i=(n_{i1}, \ldots, n_{id})^\top=\sum_{j=1}^{n_{i.}} {\boldsymbol \zeta}_{ij}, \quad i=1,2.
\end{equation}
From the independence of $X_{11}, \ldots, X_{1n_{1.}}$  and $X_{21}, \ldots, X_{2n_{2.}}$, an unbiased estimator of $\Sigma_{{\boldsymbol \pi}_{1}} \, \Sigma_{{\boldsymbol \pi}_{2}}$ is $\widehat{\Sigma}_{{\boldsymbol \pi}_{1}}\widehat{\Sigma}_{{\boldsymbol \pi}_{2}}$, and therefore, an unbiased estimator of ${\rm tr}(\Sigma_{{\boldsymbol \pi}_{1}} \, \Sigma_{{\boldsymbol \pi}_{2}})$ is ${\rm tr}(\widehat{\Sigma}_{{\boldsymbol \pi}_{1}}\widehat{\Sigma}_{{\boldsymbol \pi}_{2}})$. On the other hand, $\Sigma_{{\boldsymbol \pi}_{i}}^2$ can be unbiasedly estimated by means of
\[
\widehat{\Sigma^2}_{{\boldsymbol \pi}_{i}}  =  \frac{1}{4}\frac{1}{n_{i.}(n_{i.}-1)(n_{i.}-2)(n_{i.}-3)}\sum_{u\neq v \neq s \neq t}
({\boldsymbol \zeta}_{iu}-{\boldsymbol \zeta}_{iv})({\boldsymbol \zeta}_{iu}-{\boldsymbol \zeta}_{iv})^\top ({\boldsymbol \zeta}_{is}-{\boldsymbol \zeta}_{it})  ({\boldsymbol \zeta}_{is}-{\boldsymbol \zeta}_{it})^\top. %\\
\]
It can be easily checked that
 \[
 \widehat{\Sigma^2}_{{\boldsymbol \pi}_{i}}= \frac{1}{2}\frac{1}{n_{i.}(n_{i.}-1)}\sum_{1 \leq t \neq s \leq d} n_{it}n_{is} E_{ts} \widehat{\Sigma}_{{\boldsymbol \pi}_{i}(t,s)},
 \]
where $E_{ts}=({\boldsymbol e}_t-{\boldsymbol e}_s)({\boldsymbol e}_t-{\boldsymbol e}_s)^\top$, $\{{\boldsymbol e}_1, \ldots, {\boldsymbol e}_d\}$ is the canonical basis of $\mathbb{R}^d$ and
\begin{eqnarray*}
& & \widehat{\Sigma}_{{\boldsymbol \pi}_{i}(t,s)}=\frac{n_{i.}-2}{n_{i.}-3} \Sigma_{\hat{{\boldsymbol \pi}}_{i}(t,s)}, \quad \hat{{\boldsymbol \pi}}_{i}(t,s)=\frac{1}{n_{i.}-2}{\boldsymbol N}_i(t,s),\\ % \label{Sigma1}\\
& &  {\boldsymbol N}_i(t,s)=(n_{i1}(t,s), \ldots, n_{id}(t,s))^\top, \quad n_{ij}(t,s)=\left\{ \begin{array}{lc} n_{ij}, & j \neq t,s, \\
n_{ij}-1, & j = t,s, \end{array} \right. %\label{Sigma2}
\end{eqnarray*}
and hence,
\begin{equation} \label{tr}
{\rm tr}( \widehat{\Sigma^2}_{{\boldsymbol \pi}_{i}})= \frac{1}{2}\frac{n_{i.}-2}{n_{i.}(n_{i.}-1)(n_{i.}-3)}\sum_{1 \leq t \neq s \leq d} n_{it} n_{is}
\left\{\frac{ n_{it}+n_{is}-2}{n_{i.}-2 }-\left(\frac{ n_{it}- n_{is}}{n_{i.}-2} \right)^2 \right\}, %\, i=1,2.
\end{equation}
$i=1,2$.

Let $\widehat{var}_0(T_{U})$ denote the estimator of ${var}_0(T_{U})$ that results when  $\Sigma_{{\boldsymbol \pi}_{1r}}^2$,  $\Sigma_{{\boldsymbol \pi}_{1r}}^2$  and $\Sigma_{{\boldsymbol \pi}_{1r}} \, \Sigma_{{\boldsymbol \pi}_{2r}}$ are replaced with the estimators defined above in the expression of ${var}_0(T_{U})$.

From Corollary \ref{and0} (in the Appendix), the test that rejects $H_0$ when
$T_{U}/\sqrt{\widehat{var}_0(T_{U})}  \geq z_{1-\alpha},$
for some $\alpha \in (0,1)$,  is asymptotically correct in the sense of having asymptotic level $\alpha$.

\subsection{Test 2} \label{test.2}
Since under $H_0$ we have that $\Sigma_{{\boldsymbol \pi}_{1r}}^2=\Sigma_{{\boldsymbol \pi}_{2r}}^2=\Sigma_{{\boldsymbol \pi}_{1r}}\Sigma_{{\boldsymbol \pi}_{2r}}$, for each $r$, all these matrices can be estimated by means of $\widehat{\Sigma}_{{\boldsymbol \pi}_{1r}}\widehat{\Sigma}_{{\boldsymbol \pi}_{2r}}$. This estimator is chosen because its calculation is simpler than that of
$\widehat{\Sigma}_{{\boldsymbol \pi}_{1r}}^2$ or $\widehat{\Sigma}_{{\boldsymbol \pi}_{2r}}^2$.
Let $\widetilde{var}_0(T_{U})$ denote the estimator of ${var}_0(T_{U})$ that results when  $\Sigma_{{\boldsymbol \pi}_{1r}}^2$,  $\Sigma_{{\boldsymbol \pi}_{2r}}^2$  and $\Sigma_{{\boldsymbol \pi}_{1r}} \, \Sigma_{{\boldsymbol \pi}_{2r}}$ are all of them replaced with $\widehat{\Sigma}_{{\boldsymbol \pi}_{1r}}\widehat{\Sigma}_{{\boldsymbol \pi}_{2r}}$  in the expression of ${var}_0(T_{U})$.

From Corollary \ref{and1} (in the Appendix), the test that rejects $H_0$ when
$T_{U}/\sqrt{\widetilde{var}_0(T_{U})}  \geq z_{1-\alpha},$
for some $\alpha \in (0,1)$,  is asymptotically correct in the sense of having asymptotic level $\alpha$.

\subsection{Test 3} \label{test.3}
Since under $H_0$ we have that ${\boldsymbol \pi}_{1r}={\boldsymbol \pi}_{2r}:={\boldsymbol \pi}_r$, for each $r$,  we can  estimate
${\rm tr}(\Sigma_{{\boldsymbol \pi}_{1r}}^2)={\rm tr}(\Sigma_{{\boldsymbol \pi}_{2r}}^2)={\rm tr}(\Sigma_{{\boldsymbol \pi}_{1r}}\Sigma_{{\boldsymbol \pi}_{2r}})$ by means of the right-hand side of \eqref{tr} with the frequency vector replaced with that of the pooled sample from subpopulation $r$. Let
$\overline{var}_0(T_{U})$ denote the resulting estimator.

From Corollary \ref{and2} (in the Appendix), the test that rejects $H_0$ when
$T_{U}/\sqrt{\overline{var}_0(T_{U})}  \geq z_{1-\alpha},$
for some $\alpha \in (0,1)$,  is asymptotically correct in the sense of having asymptotic level $\alpha$.

\subsection{Power of the tests}
Section \ref{a_power} (in the Appendix) shows that, under certain not restrictive assumptions, the asymptotic power of the three tests in Subsections \ref{test.1}-- \ref{test.3} coincide. Moreover, their powers go to 1 whenever the proportion of groups where $H_0$ is not true is positive (see Theorems  \ref{power.test1}--\ref{power.test3} in the Appendix). The next section summarizes the results of a simulation study, designed to evaluate and compare their finite sample behavior.

\section{Simulation results}\label{simulation}
Three tests of $H_0$ have been proposed. To assess their finite sample performance, an extensive simulation study was carried out,  under a variety of scenarios, according to the value of $k$, the sample sizes and the number of categories $d$.

\subsection{Simulations for the level of the tests} \label{sim_level}
To analyze empirically the level, that is, the case with ${\boldsymbol \pi}_{1r}={\boldsymbol \pi}_{2r}$, $1\leq r \leq k$,  %several
two settings have been considered:
\begin{itemize} \itemsep=0pt
\item \textit{Setting 1}:
 (equiprobable case)
  ${\boldsymbol \pi}_{1r}=(1/d, \ldots,1/d)$, $\forall r$, for $d=5,10,20$. This case has been investigated in previous simulation studies involving inference research over the two-sample problem for the multinomial distribution (see \cite{SS16, matcom09}, among others).
\item \textit{Setting 2}:
        $P({\boldsymbol \pi}_{1r}={\boldsymbol \pi}^i)=0.2$, $1 \leq i \leq 5$, where ${\boldsymbol \pi}^1, \ldots , {\boldsymbol \pi}^5$ are displayed in Table \ref{tab:1},  for $d=5,10$.

\end{itemize}

\begin{table}
\centering
\begin{tabular}{cclc}
$d$ &  ${\boldsymbol \pi}^i$ %Probability vector
               & & $\|{\boldsymbol \pi}^1-{\boldsymbol \pi}^i\|^2$  \\ \hline
  5 & ${\boldsymbol \pi}^1$ & (0.2,0.2,0.2,0.2,0.2) & 0 \\
    & ${\boldsymbol \pi}^2$ & (0.1,0.15,0.2,0.25,0.3) & 0.025\\
    & ${\boldsymbol \pi}^3$ & (0.05,0.125,0.2,0.275,0.35) & 0.056\\
    & ${\boldsymbol \pi}^4$ & (0.05,0.05,0.2,0.35,0.35) & 0.090\\
    & ${\boldsymbol \pi}^5$ & (0.05,0.125,0.125,0.125,0.575) & 0.180\\   \hline
 10 & ${\boldsymbol \pi}^1$ & (0.1, 0.1, 0.1, 0.1, 0.1, 0.1, 0.1, 0.1, 0.1, 0.1) & 0\\
    & ${\boldsymbol \pi}^2$ & (0.02, 0.04, 0.06, 0.08, 0.10, 0.10, 0.12, 0.14, 0.16, 0.18) & 0.024 \\
    & ${\boldsymbol \pi}^3$ & (0.01, 0.01, 0.03, 0.03, 0.10, 0.10, 0.12, 0.2, 0.2, 0.2) & 0.056 \\
    & ${\boldsymbol \pi}^4$ & (0.01, 0.01, 0.02, 0.03, 0.03, 0.03, 0.21, 0.22, 0.22, 0.22) & 0.092 \\
    & ${\boldsymbol \pi}^5$ & (0.01, 0.01, 0.01, 0.01, 0.01, 0.01, 0.10, 0.20, 0.20, 0.44) & 0.184  \\
  \hline
\end{tabular}\caption{Values of ${\boldsymbol \pi}^1, \ldots , {\boldsymbol \pi}^5$ %Configurations of ${\boldsymbol \pi}$
                      in the Setting 2.} \label{tab:1}
\end{table}
In %all  
both settings, we performed the following experiment: $k$ pairs of random samples  were ge\-ne\-ra\-ted  from $\mathcal{M}(n_{1r.}; {\boldsymbol \pi}_{1r})$ and $\mathcal{M}(n_{2r.}; {\boldsymbol \pi}_{2r})$ with ${\boldsymbol \pi}_{1r}={\boldsymbol \pi}_{2r}$, $1\leq r \leq k$, for $k=20,50,100,200$,
 $500$,$750$;
the cases of both balanced ($n_{1r.}=n_{2r.}$, $\forall r$) and unbalanced samples, were investigated with $(n_{1r.},n_{2r.})=(5,10)$, (10,10), (20,30) and (30,30);  the $p$-values of the three tests were calculated using the normal approximation. The experiment was repeated 10,000 times, and the fraction %percentage
of $p$-values less than or equal to $\alpha=0.05$ was collected for each test. Tables \ref{tab:2}, \ref{tab:3} and \ref{tab:4} report those fractions
%percentages
for the Setting 1 and Tables \ref{tab:5} and \ref{tab:6} for the Setting 2. Since the message in all these tables is quite similar, this section only exhibits Table \ref{tab:2}. The other tables can be found in Subsection \ref{level} (in the Appendix). It is concluded from the results in these tables that:
(i) the three tests behave quite closely in terms of type I errors, and (ii) as expected from the theory, when $k$ increases, the estimated type I errors become closer to $0.05$, the nominal value, for all considered instances of $(n_{1r.},n_{2r.})$.

\begin{table}
\centering
\begin{tabular}{ccccc|ccccc}
  $k$ & $n_{1r},n_{2r}$ & Test 1 & Test 2 & Test 3 & $k$ & $n_{1r},n_{2r}$ & Test 1 & Test 2 & Test 3\\ \hline
  20  & 5,10  & 0.058 & 0.063 & 0.056 & 200 & 5,10  & 0.054 & 0.056 & 0.053 \\
      & 10,10 & 0.057 & 0.060 & 0.055 &     & 10,10 & 0.052 & 0.053 & 0.052 \\
      & 20,30 & 0.057 & 0.058 & 0.056 &     & 20,30 & 0.052 & 0.053 & 0.052 \\
      & 30,30 & 0.054 & 0.054 & 0.053 &     & 30,30 & 0.049 & 0.050 & 0.049 \\
  50  & 5,10  & 0.056 & 0.060 & 0.055 & 500 & 5,10  & 0.052 & 0.054 & 0.052 \\
      & 10,10 & 0.056 & 0.058 & 0.054 &     & 10,10 & 0.051 & 0.052 & 0.051 \\
      & 20,30 & 0.055 & 0.056 & 0.054 &     & 20,30 & 0.052 & 0.052 & 0.052 \\
      & 30,30 & 0.055 & 0.056 & 0.054 &     & 30,30 & 0.053 & 0.053 & 0.053 \\
  100 & 5,10  & 0.055 & 0.056 & 0.054 & 750 & 5,10  & 0.052 & 0.054 & 0.052 \\
      & 10,10 & 0.054 & 0.054 & 0.054 &     & 10,10 & 0.053 & 0.053 & 0.052 \\
      & 20,30 & 0.054 & 0.054 & 0.053 &     & 20,30 & 0.054 & 0.054 & 0.054 \\
      & 30,30 & 0.052 & 0.053 & 0.052 &     & 30,30 & 0.052 & 0.053 & 0.052 \\
  \hline
\end{tabular}\caption{Estimated type I error for the Setting 1 ($d=5$).} \label{tab:2}
\end{table}

An anonymous referee asked us to numerically explore other variance estimators, giving each of them a further test, namely:

\begin{enumerate}\itemsep=0pt
\item[Test 4:]corresponding to the variance estimator obtained replacing $\Sigma_{\boldsymbol{\pi}_{ir}} $ with
$\Sigma_{\hat{\boldsymbol{\pi}}_{ir}} $, in the expression of $\widehat{var}_0(T_U)$, $i=1,2$, $1 \leq r \leq k$, which is the biased analogue of the variance estimator in Test 1.

\item[Test 5:]corresponding to the variance estimator obtained replacing $\widehat{\Sigma}_{\boldsymbol{\pi}_{ir}} $ with
$\Sigma_{\hat{\boldsymbol{\pi}}_{ir}}$, in the expression of $\widetilde{var}_0(T_U)$, $i=1,2$, $1 \leq r \leq k$, which is the biased analogue of the variance estimator in Test 2.

\item[est 6:] corresponding to the variance estimator obtained 
replacing $\Sigma_{\boldsymbol{\pi}_{ir}} $ with
$\Sigma_{\hat{\boldsymbol{\pi}}_{r}} $, $i=1,2$, where $\hat{\boldsymbol{\pi}}_r$ is the pooled estimator of ${\boldsymbol{\pi}}
_r$,  $1 \leq r \leq k$, which is the biased analogue of the variance estimator in Test 3.

\item[Test 7:] corresponding to the variance estimator obtained using the nonparametric bootstrap estimator, 
that was numerically calculated by generating $B=200$ bootstrap samples, as recommended on page 47 of 
Efron and Tibshirani \cite{E&T}.
\end{enumerate}

All above variance estimators are consistent, that is,  they converge to the population variances when the sample sizes increase. To numerically study the behavior of Tests 4--7 for small to moderate sample sizes, we repeated the simulation study in Tables \ref{tab:2}, \ref{tab:3} and \ref{tab:4}  for those tests. The results obtained are displayed in Tables 
\ref{tab:level:rev1}, \ref{tab:level:rev2} and \ref{tab:level:rev3}
(in the Appendix). Looking at these tables we see that, 
for the sample sizes considered in the simulation setting, the level of the
 Tests 4--7 
does not match the target value, specially for $d=10,20$.
We also compared Tests 1-7  in terms of computing time. Table \ref{tab:cpu} (in the Appendix) displays the average time in seconds required for calculating each test statistic, for several values of $k$ and $d$ and for $n_{1r.}=n_{2r.}=30$,  $1\leq r \leq k$. On average, the time required to calculate Tests 4--6   is half than those for the proposed test. Nevertheless, looking at Table \ref{tab:cpu}, we see that time is not a problem, as the calculation is really fast. In this respect, the worst test is that based on the bootstrap variance estimator.

\subsection{Simulations for the power of the tests}\label{sim_power}

 To investigate the power of the three proposed tests, several settings have been considered. In each setting, the values of   ${\boldsymbol \pi}^1, \ldots , {\boldsymbol \pi}^5$ are as displayed in Table \ref{tab:1}:

\begin{itemize} \itemsep=0pt
\item\textit{Setting 3}: ${\boldsymbol \pi}_{1r}$ and ${\boldsymbol \pi}_{2r}$ are both randomly selected so that $P({\boldsymbol \pi}_{1r}={\boldsymbol \pi}_{2r}={\boldsymbol \pi}^i)=0.2$, $1\leq i \leq 4$ and $P({\boldsymbol \pi}_{1r}={\boldsymbol \pi}^1,\, {\boldsymbol \pi}_{2r}={\boldsymbol \pi}^0)=0.2$,
with ${\boldsymbol \pi}^0={\boldsymbol \pi}^2$ and ${\boldsymbol \pi}^0={\boldsymbol \pi}^4$, which are at an Euclidean distance of ${\boldsymbol \pi}^1$ varying from
0.025 to 0.090, for $d = 5$, and from 0.024 to 0.092 for $d = 10$, respectively. Notice that in this setting   ${\boldsymbol \pi}_{1r}$ and ${\boldsymbol \pi}_{2r}$ either coincide (with probability 0.8) or differ (with probability 0.2), and whenever they differ, they do so in exactly the same way.

\item \textit{Setting 4}: it is similar to the previous setting with  ${\boldsymbol \pi}_{1r}$ and ${\boldsymbol \pi}_{2r}$  both randomly selected so that $P({\boldsymbol \pi}_{1r}={\boldsymbol \pi}_{2r}={\boldsymbol \pi}^i)=0.2$, $1\leq i \leq 4$, but now   $P({\boldsymbol \pi}_{1r}={\boldsymbol \pi}^1,\, {\boldsymbol \pi}_{2r}={\boldsymbol \pi}^0)=0.1$ and   $P({\boldsymbol \pi}_{1r}={\boldsymbol \pi}^1,\, {\boldsymbol \pi}_{2r}={\boldsymbol \pi}^{0,rev})=0.1$, where ${\boldsymbol \pi}^{0,rev}$ is obtained from ${\boldsymbol \pi}^0$   by reversing the order of its elements,
with ${\boldsymbol \pi}^0={\boldsymbol \pi}^2$ and ${\boldsymbol \pi}^0={\boldsymbol \pi}^4$. In this setting,  ${\boldsymbol \pi}_{1r}$ and ${\boldsymbol \pi}_{2r}$ either coincide (with probability 0.8) or differ (with probability 0.2), but in contrast to the previous scenario,
half the time they are different they do it one way, and half the time they are different they do it another way.

\item\textit{Setting 5}: ${\boldsymbol \pi}_{1r}$ and ${\boldsymbol \pi}_{2r}$ are both randomly selected so that $P({\boldsymbol \pi}_{ir}={\boldsymbol \pi}^j )=0.2$, $i=1,2$, $1 \leq j \leq 5$.
 \end{itemize}

 In all settings, we performed the following experiment: $k$ pairs of random samples were generated from $\mathcal{M}(n_{1r.}; {\boldsymbol \pi}_{1r})$ and $\mathcal{M}(n_{2r.}; {\boldsymbol \pi}_{2r})$ with ${\boldsymbol \pi}_{1r}$ and ${\boldsymbol \pi}_{2r}$ selected according to each setting,  for $k=20,50, 200$, $d=5, 10$,
 and $(n_{1r.},n_{2r.})=(5,10)$, (10,10), (20,30) and (30,30); the $p$-values of the three tests were calculated using the normal approximation.
 In addition to the proposed tests, we calculated the $p$-value of the classical chi-square test (whose test statistic is as defined in (\ref{T1})) for the data without taking into account the  {division into groups} (headed in the tables as $\chi^2$), to compare the effect of the {division into groups/no division into groups} in the power of these tests. The experiment was repeated 10,000 times, and the fraction
of $p$-values less than or equal to $\alpha=0.05$ was collected for each test. Table \ref{power:1}  reports those fractions for the Setting 3, Table \ref{power:2} for the Setting 4, and Table \ref{power:3} for the Setting 5.

Looking at these tables it can be concluded  that the power of the proposed tests increases, not only according to the sample sizes and the number of populations but also depending on the magnitude of the departure from the null (the powers for ${\boldsymbol \pi}^0={\boldsymbol \pi}^4$ in Tables \ref{power:1} and \ref{power:2} are larger than the corresponding ones for ${\boldsymbol \pi}^0={\boldsymbol \pi}^2$), that is, the greater
$E(T_{U})  =  (1/\sqrt{k}) \sum_{r=1}^k \|{\boldsymbol \pi}_{1r}-{\boldsymbol \pi}_{2r}\|^2$, the larger is the power.
Additionally, with respect to the proposed tests, Test 2 exhibits in most cases a larger power than Tests 1 and 3. Finally, we discuss if it is worth considering  {division into groups}, that is, we compare the power of our proposal versus that of the chi-square test (calculated without taking into account {such division}, that is, after summing  in $r$ the vectors of frequencies $(n_{1r1}, \ldots, n_{1rd})$ and $(n_{2r1}, \ldots, n_{2rd})$). In most cases of Table \ref{power:1},  the chi-square test has a better performance than the new tests, the opposite is observed in Tables \ref{power:2} and \ref{power:3}. The main difference between the settings in those tables is that while in the Setting 3 the departure from the null hypothesis always occurs in the same fashion, in the other settings such departures may have different patterns. Similar behavior was observed in other experiments whose results are not shown in the paper.
In a practical situation, the user is rarely aware of the conceivable forms of deviations from the null. However, it seems realistic to assume that these deviations will not follow a fixed pattern, and in such a case, our proposal exhibits larger power in the scenarios considered.

\begin{table}[!h]
\centering
\begin{tabular}{ccccccccccc}
      &     &                   & \multicolumn{4}{c}{${\boldsymbol \pi}^0={\boldsymbol \pi}^2$} & \multicolumn{4}{c}{${\boldsymbol \pi}^0={\boldsymbol \pi}^4$}\\ \hline
  $d$ & $k$ & $n_{1r.}, n_{2r.}$ & Test 1 & Test 2 & Test 3  & $\chi^2$ &  Test 1 & Test 2 & Test 3  & $\chi^2$ \\  \hline
  5 & 20  & 5,5   & 0.077 & 0.082 & 0.074 & 0.061 & 0.120 & 0.130 & 0.112 & 0.120 \\
    &     & 5,10  & 0.079 & 0.087 & 0.074 & 0.073 & 0.138 & 0.147 & 0.133 & 0.171 \\
    &     & 10,10 & 0.086 & 0.091 & 0.083 & 0.086 & 0.192 & 0.199 & 0.185 & 0.214 \\
    &     & 20,30 & 0.105 & 0.107 & 0.102 & 0.121 & 0.289 & 0.295 & 0.282 & 0.325 \\
    &     & 30,30 & 0.166 & 0.169 & 0.164 & 0.176 & 0.518 & 0.585 & 0.580 & 0.500 \\
    & 50  & 5,5   & 0.080 & 0.085 & 0.076 & 0.094 & 0.150 & 0.159 & 0.145 & 0.355 \\
    &     & 5,10  & 0.084 & 0.089 & 0.081 & 0.110 & 0.193 & 0.199 & 0.186 & 0.325 \\
    &     & 10,10 & 0.097 & 0.100 & 0.096 & 0.145 & 0.302 & 0.309 & 0.298 & 0.445 \\
    &     & 20,30 & 0.132 & 0.135 & 0.130 & 0.209 & 0.470 & 0.475 & 0.465 & 0.613 \\
    &     & 30,30 & 0.242 & 0.243 & 0.240 & 0.377 & 0.856 & 0.859 & 0.856 & 0.840 \\
    & 200 & 5,5   & 0.093 & 0.095 & 0.090 & 0.242 & 0.279 & 0.286 & 0.276 & 0.754 \\
    &     & 5,10  & 0.111 & 0.115 & 0.109 & 0.336 & 0.400 & 0.407 & 0.394 & 0.869 \\
    &     & 10,10 & 0.153 & 0.155 & 0.152 & 0.482 & 0.657 & 0.661 & 0.656 & 0.960 \\
    &     & 20,30 & 0.234 & 0.237 & 0.233 & 0.672 & 0.903 & 0.904 & 0.902 & 0.992 \\
    &     & 30,30 & 0.551 & 0.552 & 0.550 & 0.926 & 0.999 & 0.999 & 0.999 & 0.999 \\ \hline
10  & 20  & 5,5   & 0.076 & 0.089 & 0.073 & 0.062 & 0.125 & 0.145 & 0.121 & 0.137 \\
    &     & 5,10  & 0.080 & 0.089 & 0.077 & 0.094 & 0.158 & 0.171 & 0.150 & 0.209 \\
    &     & 10,10 & 0.088 & 0.094 & 0.086 & 0.101 & 0.233 & 0.246 & 0.230 & 0.268 \\
    &     & 20,30 & 0.103 & 0.111 & 0.103 & 0.172 & 0.367 & 0.374 & 0.355 & 0.438 \\
    &     & 30,30 & 0.186 & 0.190 & 0.185 & 0.276 & 0.704 & 0.708 & 0.703 & 0.621 \\
    & 50  & 5,5   & 0.077 & 0.087 & 0.075 & 0.112 & 0.162 & 0.177 & 0.160 & 0.302 \\
    &     & 5,10  & 0.083 & 0.091 & 0.082 & 0.158 & 0.228 & 0.241 & 0.221 & 0.438 \\
    &     & 10,10 & 0.108 & 0.113 & 0.107 & 0.195 & 0.382 & 0.394 & 0.380 & 0.583 \\
    &     & 20,30 & 0.141 & 0.145 & 0.138 & 0.347 & 0.606 & 0.609 & 0.594 & 0.757 \\
    &     & 30,30 & 0.298 & 0.302 & 0.296 & 0.564 & 0.936 & 0.937 & 0.937 & 0.921 \\
    & 200 & 5,5   & 0.102 & 0.109 & 0.101 & 0.386 & 0.372 & 0.385 & 0.372 & 0.909 \\
    &     & 5,10  & 0.123 & 0.127 & 0.121 & 0.544 & 0.528 & 0.534 & 0.518 & 0.969 \\
    &     & 10,10 & 0.175 & 0.179 & 0.175 & 0.725 & 0.815 & 0.820 & 0.815 & 0.996 \\
    &     & 20,30 & 0.277 & 0.280 & 0.273 & 0.907 & 0.969 & 0.976 & 0.975 & 0.999 \\
    &     & 30,30 & 0.664 & 0.666 & 0.664 & 0.994 & 1     & 1     & 1     & 1     \\ \hline
\end{tabular}\caption{Estimated power for  the Setting 3.}\label{power:1}
\end{table}

\begin{table}[!h]
\centering
\begin{tabular}{ccccccccccc}
      &     &                   & \multicolumn{4}{c}{${\boldsymbol \pi}^0={\boldsymbol \pi}^2$} & \multicolumn{4}{c}{${\boldsymbol \pi}^0={\boldsymbol \pi}^4$}\\ \hline
  $d$ & $k$ & $n_{1r.}, n_{2r.}$ & Test 1 & Test 2 & Test 3  & $\chi^2$ &  Test 1 & Test 2 & Test 3  & $\chi^2$ \\  \hline
  5 & 20  & 5,5   & 0.077 & 0.083 & 0.070 & 0.047 & 0.097 & 0.105 & 0.089 & 0.057 \\
    &     & 5,10  & 0.078 & 0.084 & 0.073 & 0.050 & 0.105 & 0.112 & 0.100 & 0.058 \\
    &     & 10,10 & 0.087 & 0.092 & 0.084 & 0.050 & 0.140 & 0.146 & 0.137 & 0.068 \\
    &     & 20,30 & 0.105 & 0.109 & 0.102 & 0.056 & 0.181 & 0.186 & 0.177 & 0.076 \\
    &     & 30,30 & 0.166 & 0.170 & 0.166 & 0.069 & 0.385 & 0.389 & 0.383 & 0.129 \\
    & 50  & 5,5   & 0.079 & 0.083 & 0.075 & 0.050 & 0.109 & 0.115 & 0.104 & 0.060 \\
    &     & 5,10  & 0.079 & 0.084 & 0.077 & 0.052 & 0.126 & 0.133 & 0.123 & 0.066 \\
    &     & 10,10 & 0.102 & 0.105 & 0.101 & 0.052 & 0.182 & 0.186 & 0.179 & 0.075 \\
    &     & 20,30 & 0.131 & 0.133 & 0.128 & 0.059 & 0.286 & 0.289 & 0.285 & 0.099 \\
    &     & 30,30 & 0.256 & 0.247 & 0.244 & 0.067 & 0.614 & 0.615 & 0.613 & 0.168 \\
    & 200 & 5,5   & 0.097 & 0.105 & 0.095 & 0.046 & 0.173 & 0.179 & 0.171 & 0.086 \\
    &     & 5,10  & 0.113 & 0.116 & 0.111 & 0.050 & 0.230 & 0.235 & 0.226 & 0.102 \\
    &     & 10,10 & 0.151 & 0.153 & 0.148 & 0.055 & 0.390 & 0.394 & 0.388 & 0.141 \\
    &     & 20,30 & 0.230 & 0.238 & 0.228 & 0.069 & 0.630 & 0.632 & 0.627 & 0.205 \\
    &     & 30,30 & 0.548 & 0.550 & 0.548 & 0.069 & 0.971 & 0.972 & 0.972 & 0.365 \\ \hline
10  & 20  & 5,5   & 0.096 & 0.102 & 0.090 & 0.056 & 0.094 & 0.110 & 0.090 & 0.053 \\
    &     & 5,10  & 0.111 & 0.126 & 0.108 & 0.064 & 0.104 & 0.127 & 0.124 & 0.062 \\
    &     & 10,10 & 0.158 & 0.170 & 0.156 & 0.083 & 0.157 & 0.166 & 0.154 & 0.080 \\
    &     & 20,30 & 0.225 & 0.233 & 0.219 & 0.103 & 0.219 & 0.226 & 0.213 & 0.105 \\
    &     & 30,30 & 0.471 & 0.477 & 0.471 & 0.183 & 0.464 & 0.469 & 0.463 & 0.191 \\
    & 50  & 5,5   & 0.123 & 0.134 & 0.121 & 0.071 & 0.115 & 0.126 & 0.112 & 0.073 \\
    &     & 5,10  & 0.142 & 0.151 & 0.137 & 0.080 & 0.153 & 0.165 & 0.148 & 0.083 \\
    &     & 10,10 & 0.230 & 0.240 & 0.228 & 0.116 & 0.230 & 0.239 & 0.228 & 0.109 \\
    &     & 20,30 & 0.362 & 0.367 & 0.353 & 0.156 & 0.364 & 0.368 & 0.354 & 0.151 \\
    &     & 30,30 & 0.744 & 0.746 & 0.743 & 0.290 & 0.736 & 0.739 & 0.736 & 0.295 \\
    & 200 & 5,5   & 0.204 & 0.214 & 0.204 & 0.142 & 0.208 & 0.217 & 0.206 & 0.145 \\
    &     & 5,10  & 0.296 & 0.304 & 0.290 & 0.184 & 0.292 & 0.298 & 0.285 & 0.186 \\
    &     & 10,10 & 0.512 & 0.520 & 0.512 & 0.283 & 0.517 & 0.523 & 0.516 & 0.286 \\
    &     & 20,30 & 0.785 & 0.785 & 0.777 & 0.415 & 0.780 & 0.780 & 0.773 & 0.412 \\
    &     & 30,30 & 0.996 & 0.996 & 0.996 & 0.682 & 0.994 & 0.995 & 0.995 & 0.680 \\ \hline
\end{tabular}\caption{Estimated power for the Setting 4.}\label{power:2}
\end{table}

\begin{table}[!h]
\centering
\begin{tabular}{ccccccccccc}
 & & \multicolumn{4}{c}{$d=5$} & &  \multicolumn{4}{c}{$d=10$}\\ \hline
  $k$ & $n_{1r.}, n_{2r.}$ & Test 1 & Test 2 & Test 3  & $\chi^2$ & & Test 1 & Test 2 & Test 3  & $\chi^2$\\  \hline
 20  &  5,5  & 0.293 & 0.311 & 0.281 & 0.077  & &  0.320 & 0.347 & 0.312 & 0.082 \\
     &  5,10 & 0.393 & 0.404 & 0.383 & 0.092  & &  0.436 & 0.460 & 0.432 & 0.102 \\
     & 10,10 & 0.609 & 0.617 & 0.602 & 0.120  & &  0.689 & 0.703 & 0.685 & 0.144 \\
     & 20,30 & 0.820 & 0.823 & 0.818 & 0.156  & &  0.889 & 0.895 & 0.888 & 0.213 \\
     & 30,30 & 0.990 & 0.990 & 0.990 & 0.290  & &  0.997 & 0.997 & 0.997 & 0.372 \\
 50  &  5,5  & 0.486 & 0.496 & 0.473 & 0.077  & &  0.545 & 0.564 & 0.540 & 0.084 \\
     &  5,10 & 0.638 & 0.645 & 0.631 & 0.094  & &  0.720 & 0.738 & 0.718 & 0.107 \\
     & 10,10 & 0.891 & 0.894 & 0.888 & 0.121  & &  0.731 & 0.756 & 0.729 & 0.149 \\
     & 20,30 & 0.985 & 0.986 & 0.985 & 0.159  & &  0.787 & 0.793 & 0.789 & 0.216 \\
     & 30,30 & 1     & 1     & 1     & 0.294  & &  0.862 & 0.872 & 0.860 & 0.367 \\
 200 &  5,5  & 0.909 & 0.919 & 0.907 & 0.078  & &  0.956 & 0.958 & 0.956 & 0.094 \\
     &  5,10 & 0.952 & 0.962 & 0.948 & 0.095  & &  0.993 & 0.993 & 0.993 & 0.109 \\
     & 10,10 & 0.999 & 0.999 & 0.999 & 0.123  & &  0.996 & 0.996 & 0.996 & 0.155 \\
     & 20,30 & 1     & 1     & 1     & 0.164  & &  0.998 & 0.999 & 0.998 & 0.284 \\
     & 30,30 & 1     & 1     & 1     & 0.300 &  &  0.999 & 0.999 & 0.999 & 0.364 \\ \hline
\end{tabular}\caption{Estimated power for the Setting 5.}\label{power:3}
\end{table}

 To test $H_0$, one can also apply a test of each of the hypothesis that composes $H_0$, that is, a test of $H_{0r}:\, {\boldsymbol \pi}_{1r}={\boldsymbol \pi}_{2r}$ against  $H_{1r}:\, {\boldsymbol \pi}_{1r}\neq {\boldsymbol \pi}_{2r}$, $1 \leq r \leq k$, and then apply
either the Bonferroni method, which controls the family-wise error rate, or the Benjamini-Hochberg method (see \cite{Benjamini2001}), which controls the false discovery rate when the $k$ tests are
independent. Both procedures agree in rejecting $H_0$ if
$\min_{1 \leq i \leq k}p_i \leq \alpha/k$, where $p_1, \ldots, p_k$ are the $p$-values obtained when testing  $H_{01}, \ldots,  H_{0k}$, respectively. Since we are dealing with small sample sizes, some cell frequencies can be equal to zero, and this fact prevents us from using the $\chi^2$ test for testing $H_{0r}$, $1 \leq r \leq k$. Instead, we used the test statistic in \eqref{TU1}. To calculate $p_1, \ldots, p_k$ one can approximate the null distribution of the test statistic employing either an estimation of its asymptotic null distribution or a bootstrap estimator. The results  in \cite{matcom09}, for the comparison of multinomial populations, and in \cite{IS2010,Statistics2014}, for some goodness-of-fit problems in multinomial populations, indicate that
 it is worth
calculating the bootstrap estimator, because it is more accurate than the approximation yielded by the asymptotic null distribution. To obtain the bootstrap estimator of the $p$-value for testing $H_{0r}$,  we first compute  the value of the test statistic at the observed data ${\boldsymbol N}_{1r}$ and ${\boldsymbol N}_{2r}$. Recall that under the null hypothesis $ {\boldsymbol \pi}_{1r}= {\boldsymbol \pi}_{2r}:= {\boldsymbol \pi}_{r}$, and thus
\[
{\boldsymbol N}_{1r} \overset{H_{0r}}{\sim}   \mathcal{M}(n_{1r.}; {\boldsymbol \pi}_{r}),\quad
{\boldsymbol N}_{2r} \overset{H_{0r}}{\sim}   \mathcal{M}(n_{2r.}; {\boldsymbol \pi}_{r}).
\]
Then,  $B=1000$ bootstrap replications of the test statistic are obtained by generating
\[
{\boldsymbol N}_{1r}^{*b}  {\sim}   \mathcal{M}(n_{1r.}; \hat{\boldsymbol \pi}_{r}), \quad
{\boldsymbol N}_{2r}^{*b}  {\sim}   \mathcal{M}(n_{2r.}; \hat{\boldsymbol \pi}_{r}), \quad 1\leq b\leq B,
\]
where $\hat{\boldsymbol \pi}_{r}=({\boldsymbol N}_{1r}+{\boldsymbol N}_{2r})/(n_{1r.}+n_{2r.})$, and computing the test statistic at
 ${\boldsymbol N}_{1r}^{*b}$ and ${\boldsymbol N}_{2r}^{*b}$, $1\leq b\leq B$. The bootstrap estimator of the $p$-value is the proportion of bootstrap replications that are greater than the observed value of the test statistic. Clearly, this procedure is much more time-consuming than our proposal, which uses the percentiles of the standard normal distribution. Because of this reason, the powers in Table
 \ref{power:CM} are based on 1,000 repetitions.  Table
 \ref{power:CM} reports the estimated powers in the Settings 3--5 using the described procedure. Comparing the results in Tables \ref{power:1}--\ref{power:3} with those in Table \ref{power:CM}, one can see that in most cases, the new proposal is more powerful than the one based on multiple tests.

\begin{table}[!h]
\centering
\begin{tabular}{ccccccccccccc}
 & & \multicolumn{4}{c}{$d=5$} & &  \multicolumn{4}{c}{$d=10$}\\ \hline
  $k$ & $n_{1r.}, n_{2r.}$ & (3.1) & (3.2) & (4.1) & (4.2) & (5)  & &   (3.1) & (3.2) & (4.1) & (4.2) & (5)  \\ \hline
 20  &  5,5  & 0.028 & 0.029 & 0.027 & 0.043 & 0.064 & & 0.014 & 0.009 & 0.011 & 0.009 & 0.024  \\
     &  5,10 & 0.035 & 0.055 & 0.035 & 0.057 & 0.121 & & 0.026 & 0.027 & 0.027 & 0.027 & 0.086 \\
     & 10,10 & 0.046 & 0.088 & 0.040 & 0.100 & 0.239 & & 0.028 & 0.045 & 0.029 & 0.073 & 0.171 \\
     & 20,30 & 0.081 & 0.342 & 0.101 & 0.332 & 0.754 & & 0.071 & 0.447 & 0.075 & 0.427 & 0.762\\
     & 30,30 & 0.117 & 0.493 & 0.131 & 0.506 & 0.838 & & 0.106 & 0.658 & 0.135 & 0.675 & 0.907\\
 50  &  5,5  & 0.012 & 0.009 & 0.009 & 0.008 & 0.024 & & 0.005 & 0.004 & 0.003 & 0.002 & 0.008 \\
     &  5,10 & 0.042 & 0.043 & 0.025 & 0.052 & 0.102 & & 0.013 & 0.019 & 0.012 & 0.015 & 0.048 \\
     & 10,10 & 0.033 & 0.082 & 0.040 & 0.074 & 0.218 & & 0.017 & 0.040 & 0.014 & 0.042 & 0.160 \\
     & 20,30 & 0.088 & 0.361 & 0.084 & 0.358 & 0.819 & & 0.070 & 0.420 & 0.061 & 0.462 & 0.874 \\
     & 30,30 & 0.115 & 0.539 & 0.106 & 0.544 & 0.923 & & 0.077 & 0.737 & 0.098 & 0.717 & 0.968 \\
 200 &  5,5  & 0.030 & 0.044 & 0.038 & 0.050 & 0.089 & & 0.006 & 0.013 & 0.005 & 0.010 & 0.040 \\
     &  5,10 & 0.109 & 0.171 & 0.126 & 0.150 & 0.394 & & 0.058 & 0.082 & 0.054 & 0.064 & 0.248 \\
     & 10,10 & 0.151 & 0.260 & 0.161 & 0.289 & 0.657 & & 0.067 & 0.130 & 0.058 & 0.125 & 0.492 \\
     & 20,30 & 0.284 & 0.833 & 0.305 & 0.839 & 0.998 & & 0.201 & 0.886 & 0.227 & 0.899 & 1   \\
     & 30,30 & 0.333 & 0.958 & 0.359 & 0.959 & 1     & & 0.310 & 0.999 & 0.319 & 0.996 & 1    \\ \hline
\end{tabular}\caption{Estimated power using multiple tests for:
(3.1) the Setting 3 with ${\boldsymbol \pi}^0={\boldsymbol \pi}^2$,
(3.2) the Setting 3 with ${\boldsymbol \pi}^0={\boldsymbol \pi}^4$,
(4.1) the Setting 4 with ${\boldsymbol \pi}^0={\boldsymbol \pi}^2$,
(4.2) the Setting 4 with ${\boldsymbol \pi}^0={\boldsymbol \pi}^4$,
(5) the Setting 5.}\label{power:CM}
\end{table}

\section{Application to three data sets}\label{application}
This section presents three examples of the application of the tests proposed in Section \ref{The.test.statistic}
 in very diverse areas of science such as education, COVID-19 pandemic and digital
elevation models.

\subsection{Education level}
 The considered data set contains complete population information of 49,563 individuals. It is available in the \texttt{NonProbEst} R-package \cite{RJ-2020-015}. For each individual, age, gender and education level achieved (understood as Primary, Secondary and Tertiary Education)  were collected. The objective is simultaneously testing whether there exists a significant difference in the education level achieved by men and women in each of $k=71$ groups defined by ages, which range between 20 and 90.
With this aim, independent random samples were drawn from each population (men and women) inside each group (age), with sample sizes randomly chosen from 4 to 6 for each age and gender group. Tests 1-3 were applied and the corresponding $p$-values are 0.0170, 0.01510 and 0.0196. Therefore, it can be concluded that there are significant differences in the education level between men and women when age is incorporated to define subgroups. In addition, the chi-square test was also applied to test if the population proportion of each education level is the same in men and women (without taking into account the division by age). The sample sizes corresponding to men and women (without groups) are 361 and 353, respectively, which are large enough to ensure the validity of the chi-squared approximation to the null distribution of the $\chi^2$ test statistic.  The  $p$-value of the chi-square is 0.5249. So, in this case, the use of the chi-square test fades the reality in the sense that not for all ages, men and women have reached the same education level.

We also tested each of the $k=71$ null hypotheses $H_{0r}:\, {\boldsymbol \pi}_{1r}={\boldsymbol \pi}_{2r}$ against  $H_{1r}:\, {\boldsymbol \pi}_{1r}\neq {\boldsymbol \pi}_{2r}$, $1 \leq r \leq k$, as described in Subsection \ref{sim_power}.
 After obtaining the bootstrap $p$-values for each test, we adjusted them using the Benjamini-Hochberg method (which is less stringent than methods that control the error rate by families). No hypothesis was rejected.

\subsection{COVID data}

We also considered the data set that contains the responses of the ESPACOV II poll \cite{espacov-2}, that was carried out in Spain during the COVID-19 pandemic from January 18 to 25, 2021. At that time, the vaccination campaign
in Spain had just started in nursing homes and the country was in the midst of the third wave of infections, the level of information had increased, but the population continued to have limited social relationships.
The selected sample can be considered as independent and identically distributed observations (see \cite{espacov-2} for details). The study contains 101 recorded variables for a total of 1,644 citizens. Here, we study the following variables:
\begin{itemize} \itemsep=0pt
\item The provinces, that are used to define the groups. In each case, only those provinces (among 52) for which the sample sizes $n_{1r.}$ and $n_{2r.}$  are greater than or equal to 5 are considered.
 \item Gender, with outcomes male and female, will be used to define the two populations to be compared in each group.
 \item ICL, that corresponds to the question: Has anyone in your close relatives been infected with coronavirus?. It has two outcomes: yes and no,  that will be used to define the two populations to be compared in each group.
 \item INF,  that corresponds to the question: Since the pandemic began, have you been infected with coronavirus?. It has five outcomes:
    No, I got infected and I'm cured, I got infected and I still have sequels,
I think I'm infected but without proof, and I am infected. It is used as a target variable to be compared.
 \item EFF,  that corresponds to the question: Do you consider yourself informed about the effectiveness of vaccines?. It has five outcomes:
   No informed, Little informed, Somewhat informed, Fairly informed and Very informed.  It is used as a target variable to be compared.
 \item SEC, that corresponds to the question: Do you consider yourself informed about the secondary effects of vaccines?. It has five outcomes:
   No informed, Little informed, Somewhat informed, Fairly informed and Very informed.  It is used as a target variable to be compared.
\end{itemize}

The results of the application to each case of the three tests proposed are displayed in Table \ref{caso2}. Besides, the chi-square test and the sample sizes considered in its application are also included. From these results, we can highlight that: there are significant differences in the degree of infection by COVID-19 between men and women, inside the provinces but not at the country level; the same can be concluded for the information on the the effectiveness of the vaccine among citizens with relatives infected/not infected; however, we cannot conclude differences, neither by provinces nor at the country level, among citizens with relatives infected/not infected, about the information on the secondary effects of the vaccine.

In each case, we also tested the  null hypotheses $H_{0r}:\, {\boldsymbol \pi}_{1r}={\boldsymbol \pi}_{2r}$ against  $H_{1r}:\, {\boldsymbol \pi}_{1r}\neq {\boldsymbol \pi}_{2r}$, $1 \leq r \leq k$, as described in Subsection \ref{sim_power}.
 After obtaining the bootstrap $p$-values for each test, we adjusted them using the Benjamini-Hochberg method. No hypothesis was rejected.

\begin{table}
\begin{tabular}{cccccccc}
    & \multicolumn{4}{c}{$p$-values} &  &  \\ \hline
  \begin{tabular}{c}Target \\Variable \end{tabular} & $T_1$ & $T_2$ & $T_3$ & $\chi^2$ & \begin{tabular}{c} Variable\\ defining the \\ two populations \end{tabular} & $k$ & $n_{1..},n_{2..}$ \\ \hline
  INF & 0.0476 & 0.0163 & 0.0877 & 0.3537 & Gender & 39 & Men (711), Women (865) \\
  EFF & 0.0660 & 0.0524 & 0.0726 & 0.2954 & ICL & 34 & Yes (454), No (1005) \\
  SEC & 0.6925 & 0.6741 & 0.6838 & 0.7976 & ICL & 34 & Yes (454), No (1005)\\
  \hline
\end{tabular}\caption{Application to ESPACOV II study.}\label{caso2}
\end{table}

\subsection{Digital elevation models}

A Digital Elevation Model (DEM) is a bare earth elevation model representing the surface of the Earth, which means elevations of the terrain (bare Earth) void of vegetation and man-made features. DEMs allow us to directly measure elevations and are the basis for other models, such as slopes, orientations (aspect), insolation, drainage networks, visual and watershed analysis, etc., which can be easily derived from them through specific algorithms. To date, the comparison of two DEMs covering the same area has been tackled by analyzing the discrepancy between the elevations of the two models. Here, we suggest a new approach to make such a comparison based on a categorical variable obtained from a morphological parameter of the land.

\begin{figure}
  \centering
  \includegraphics[width=7cm]{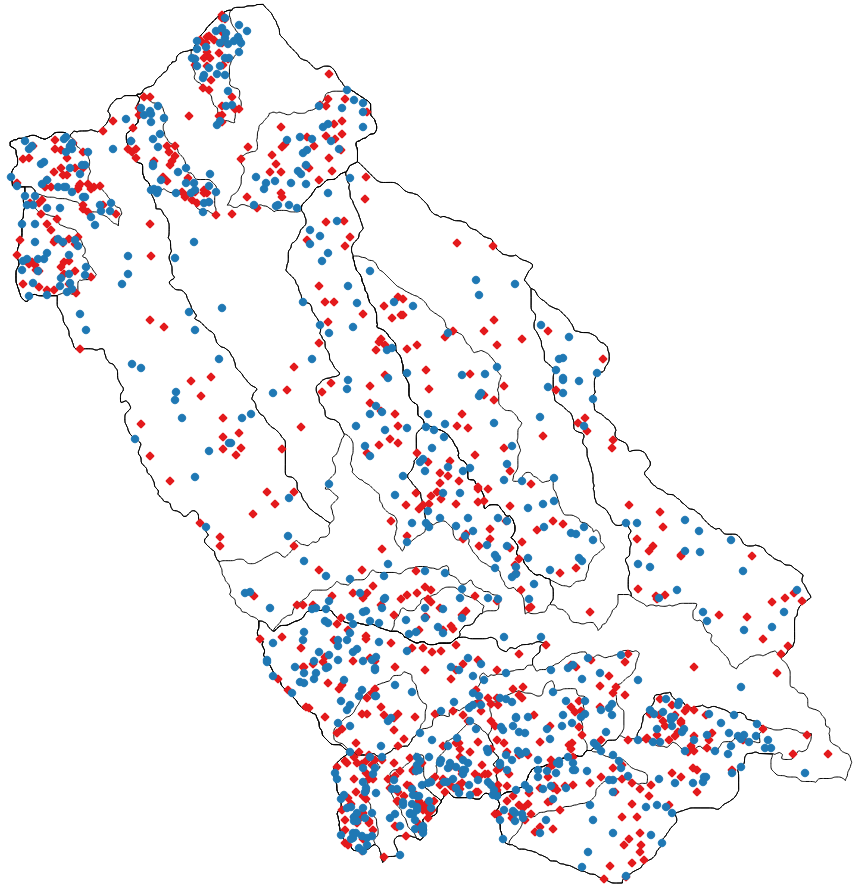} \\
  \caption{Samples of points per watershed in red for ASTER and in green for MDT25 ($n_{1r} =
n_{2r} = 30$).}\label{fig:sample}
\end{figure}

To this end, let us consider two gridded DEMs over the Allo area (Navarra, Spain): \textsc{MDT25} and \textsc{ASTER} (see for details https://pnoa.ign.es/estado-del-proyecto-lidar/primera-cobertura, https://asterweb.jpl.nasa.gov/gdem.asp). To carry  out the comparison, the terrains were classified into $d=10$ categories of landforms, derived from the topographic position index (TPI) (\cite{Weiss01}), that was observed in a total of  $k=22$ watersheds in the study area. From each watershed, two sets of independent random samples of $n_{1r.}=n_{2r.}=30$ points were obtained, the target variable was observed at such points, and the procedure proposed in Section \ref{The.test.statistic} applied. Figure \ref{fig:sample} displays such samples. Recall from the Introduction, that independence is assumed to hold in this setting. However, this kind of data may be affected by spatial correlation and users must check this assumption before applying this proposal. In our case, this point has been evaluated by simulation and the categorical variable considered to compare two DEMs does not seem to be affected by such correlation.

The application of Tests 1-3  to the sampled data gave the following $p$-values: 0.477, 0.477 and 0.478, respectively. When data is compared without taking into account the watersheds (groups) using the $\chi^2$ test, the sample size was 660 in each DEM, and the $p$-value is 0.168. Therefore, when using the TPI index to evaluate the homogeneity between ASTER and MDT25 DEMs through the spatialization induced by watersheds, it can be concluded that there is no evidence against the equality of the two DEMs. The same conclusion is achieved when such spatialization is not taken into account.

We also tested each of the $k=22$ null hypotheses $H_{0r}:\, {\boldsymbol \pi}_{1r}={\boldsymbol \pi}_{2r}$ against  $H_{1r}:\, {\boldsymbol \pi}_{1r}\neq {\boldsymbol \pi}_{2r}$, $1 \leq r \leq k$, as described in Subsection \ref{sim_power}.
 After obtaining the bootstrap p-values for each test, we adjusted them using the Benjamini-Hochberg method. No hypothesis was rejected.

\section{Conclusions}  \label{conclusions}

This paper deals with the two-sample problem for categorical data when each population is divided into subpopulations or groups. In such a case, the distribution of the target variable may vary across subpopulations, moreover, the two categorical variables may have the same distribution in the whole population, but their distributions could differ in some groups. So, instead of testing the homogeneity of the two categorical variables, one may be interested in simultaneously testing the homogeneity in all groups. A procedure for carrying out that testing problem has been proposed and analysed, both theoretically and numerically.

Moreover, the proposal is very easy to implement,  not requiring the use of complicated resampling
methods to get $p$-values. It applies whenever the sample sizes are comparable,  allowing them to remain bounded or to increase with $k$. The latter point is a remarkable aspect of the proposal because in practice, when a large number of groups is built, it is not realistic to expect to have a large number of data in each subgroup (or subpopulation).

\section{Appendix} \label{appendix}
All theoretical derivations are presented here, as well as some further simulation results. Subsection \ref{UIT} obtains the UIT of $H_0$ vs $H_1$.  Subsection \ref{level:LRT} displays some simulation results for the LRT. Subsection \ref{u_est} derives an unbiased estimator of $\|{\boldsymbol \pi}_1-{\boldsymbol \pi}_2\|^2$. Subsection \ref{Asymptotic} calculates the asymptotic distribution of $T_U$ and gives properties of the variance estimators defined in Section \ref{The.test.statistic}.  Subsection \ref{a_power} studies the asymptotic power of the test associated with each variance estimator. The proofs of the results in Subsections \ref{Asymptotic} and \ref{a_power}  can be found in Subsection \ref{proofs}. Subsection \ref{level} displays some tables with numerical results for the level, that were described in Subsection \ref{sim_level}. 

\subsection{The union-intersection test} \label{UIT}
This subsection derives the UIT of $H_0$ vs $H_1$.
 With this aim, we closely follow the reasoning in \cite{Anderson}, which calculated the UIT for each of the hypotheses $H_{0r}:  \, {\boldsymbol \pi}_{1r}={\boldsymbol \pi}_{2r}$ vs
$H_{1r}:  \, {\boldsymbol \pi}_{1r} \neq {\boldsymbol \pi}_{2r}$,  $1 \leq r \leq k$.

As  $\sum_{j=1}^d \pi_{irj}=1$, $i=1,2$,  $1\leq  r \leq k$, $H_0$  is equivalent to 
\begin{eqnarray}
H_0:  & {\boldsymbol \varpi}_{1r}={\boldsymbol \varpi}_{2r},& 1 \leq r \leq k, \label{otra}
\end{eqnarray}
where ${\boldsymbol \varpi}_{ir}=(\pi_{ir1}, \ldots, \pi_{ird-1})^\top$, $i=1,2$, $1\leq  i \leq k$. On the other hand, the null hypothesis in \eqref{otra} is true if and only if
\begin{eqnarray*}
H_{0, {\boldsymbol a} }:  & 
{\boldsymbol a}^\top({\boldsymbol \varpi}_1-{\boldsymbol \varpi}_2)=0 
\end{eqnarray*}
is true $\forall {\boldsymbol a}=({\boldsymbol a}_1^\top , \ldots, {\boldsymbol a}_k^\top )^\top \in \mathbb{R}^{k(d-1)} \setminus \{0\}$, where ${\boldsymbol \varpi}_i=({\boldsymbol \varpi}_{i1}^\top, \ldots, {\boldsymbol \varpi}_{ik}^\top)^\top$,  $i=1,2$, and ${\boldsymbol a}_r \in \mathbb{R}^{d-1}$, $1 \leq r \leq k$.

As $\widehat{{\boldsymbol \varpi}}_{ir}=(n_{ir1}, \ldots, n_{ird-1})^\top/n_{ir.}$ is an unbiased estimator of ${{\boldsymbol \varpi}}_{ir}$, $i=1,2$, $1\leq r \leq k$, we have that
 ${\boldsymbol a}^\top(\widehat{\boldsymbol \varpi}_1-\widehat{\boldsymbol \varpi}_{2})$ is unbiased for
$ {\boldsymbol a}^\top({\boldsymbol \varpi}_1-{\boldsymbol \varpi}_{2})$, where $\widehat{\boldsymbol \varpi}_i=(\widehat{\boldsymbol \varpi}_{i1}^\top, \ldots, \widehat{\boldsymbol \varpi}_{ik}^\top)^\top$, $i=1,2$, with variance 

$\sigma^2_{ {\boldsymbol a}}=\sum_{r=1}^k {\boldsymbol a}_i^\top (\Sigma_{{\boldsymbol \varpi}_{1r}}/n_{1r.}+ \Sigma_{{\boldsymbol \varpi}_{2r}}/n_{2r.}) {\boldsymbol a}_i$.

Under $H_{0r}$, $\Sigma_{{\boldsymbol \varpi}_{1r}}=\Sigma_{{\boldsymbol \varpi}_{2r}}:=\Sigma_r$. Let $\widehat{\Sigma}_r$ denote the maximum likelihood estimator of $\Sigma_r$ and let  $\hat{\sigma}^2_{ {\boldsymbol a}}=\sum_{r=1}^k (1/n_{1r.}+ 1/n_{2r.}) {\boldsymbol a}_i^\top \widehat{\Sigma}_r {\boldsymbol a}_i$.

It is reasonable to reject 
$H_{0, {\boldsymbol a} }$ for large values of 
$${\cal T}_{\boldsymbol a} = \left \{ {\boldsymbol a}^\top(\widehat{\boldsymbol \varpi}_1-\widehat{\boldsymbol \varpi}_{2})\right \}^2/ \hat{\sigma}^2_{ {\boldsymbol a}}.$$
However, we are not interested in testing $H_{0, {\boldsymbol a} }$ for a particular ${\boldsymbol a} $, but for all   ${\boldsymbol a} \neq 0$. According to the union-intersection principle,  the test statistic is
$${\cal T}=\sup_{{\boldsymbol a} \neq 0} {\cal T}_{\boldsymbol a}.
$$
${\cal T}$ is the largest eigenvalue of the matrix ${\cal M}=(\widehat{\boldsymbol \varpi}_1-\widehat{\boldsymbol \varpi}_{2})(\widehat{\boldsymbol \varpi}_1-\widehat{\boldsymbol \varpi}_{2})^\top \widehat{\Sigma}^{-1}$, where $\widehat{\Sigma}$ is the block diagonal matrix with diagonal blocks  $(1/n_{11.}+ 1/n_{21.})\widehat{\Sigma}_{1}, \ldots, (1/n_{1k.}+ 1/n_{2k.})\widehat{\Sigma}_{k}$ (see, e.g. \cite{Seber}). As  ${\cal M}$ has range one, it readily follows that
${\cal T}=(\widehat{\boldsymbol \varpi}_1-\widehat{\boldsymbol \varpi}_2)^\top \hat{\Sigma}^{-1}
(\widehat{\boldsymbol \varpi}_1-\widehat{\boldsymbol \varpi}_{2})$. With the help of the computations made in \cite{Anderson}, it follows that ${\cal T}=\sum_{r=1}^k T_r^2$, where 
$T_{r}$ denotes the  chi-square test statistic for testing $H_{0r}$ .

\subsection{ Simulation results for the LRT} \label{level:LRT}

Tables \ref{tab:TRV1} and \ref{tab:TRV2} are the analogues of Tables \ref{tab:8} and \ref{tab:88} obtained for the LRT.

\begin{table}[ht!]
\centering
\begin{tabular}{ccccc|ccccc}

      & &\multicolumn{3}{c}{$d$} & & & \multicolumn{3}{c}{$d$} \\    \hline
  $k$ & $n_{1r},n_{2r}$ & 5 & 10 & 20 & $k$ & $n_{1r},n_{2r}$ & 5 & 10 & 20 \\ \hline
  20  & 5,10  & 0.735 & 0.226 &  0    & 200 & 5,10  & 1.000 & 0.813 & 0 \\
      & 10,10 & 0.665 & 0.824 &  0    &     & 10,10 & 1.000 & 1.000 & 0   \\
      & 20,30 & 0.281 & 0.760 & 0.968 &     & 20,30 & 0.714 & 1.000 & 1.000  \\
      & 30,30 & 0.228 & 0.624 & 0.982 &     & 30,30 & 0.542 & 0.997 & 1.000  \\
  50  & 5,10  & 0.958 & 0.379 &  0    & 500 & 5,10  & 1.000 & 0.951 & 0 \\
      & 10,10 & 0.908 & 0.985 &  0    &     & 10,10 & 1.000 & 1.000 & 0 \\
      & 20,30 & 0.395 & 0.961 & 0.999 &     & 20,30 & 0.936 & 1.000 & 1.000   \\
      & 30,30 & 0.297 & 0.875 & 0.999 &     & 30,30 & 0.784 & 1.000 & 1.000  \\
  100 & 5,10  & 0.993 & 0.574 &  0    & 750 & 5,10  & 1.000 & 0.993 & 0  \\
      & 10,10 & 0.990 & 1.000 &  0    &     & 10,10 & 1.000 & 1.000 & 0 \\
      & 20,30 & 0.530 & 0.994 & 1.000 &     & 20,30 & 0.954 & 1.000 & 1.000 \\
      & 30,30 & 0.392 & 0.977 & 1.000 &     & 30,30 & 0.889 & 1.000 & 1.000 \\
  \hline
\end{tabular}\caption{Estimated type I error for the test with critical region $V_k>z_{1-\alpha}$.}
\label{tab:TRV1}
\end{table}

\begin{table}[ht!]
\centering
\begin{tabular}{ccccc|ccccc}

      & &\multicolumn{3}{c}{$d$} & & & \multicolumn{3}{c}{$d$} \\    \hline
  $k$ & $n_{1r},n_{2r}$ & 5 & 10 & 20 & $k$ & $n_{1r},n_{2r}$ & 5 & 10 & 20 \\ \hline
  20  & 5,10  & 0.054 & 0.050 & 0.048 & 200 & 5,10  & 0.053 & 0.050 & 0.050\\
      & 10,10 & 0.055 & 0.052 & 0.053 &     & 10,10 & 0.047 & 0.049 & 0.050  \\
      & 20,30 & 0.056 & 0.054 & 0.051 &     & 20,30 & 0.049 & 0.054 & 0.051 \\
      & 30,30 & 0.058 & 0.049 & 0.056 &     & 30,30 & 0.056 & 0.054 & 0.049 \\
  50  & 5,10  & 0.049 & 0.048 & 0.048 & 500 & 5,10  & 0.048 & 0.048 & 0.046\\
      & 10,10 & 0.051 & 0.050 & 0.049 &     & 10,10 & 0.051 & 0.050 & 0.047\\
      & 20,30 & 0.055 & 0.052 & 0.051 &     & 20,30 & 0.050 & 0.054 & 0.052  \\
      & 30,30 & 0.056 & 0.053 & 0.051 &     & 30,30 & 0.054 & 0.051 & 0.047 \\
  100 & 5,10  & 0.053 & 0.052 & 0.054 & 750 & 5,10  & 0.053 & 0.048 & 0.047 \\
      & 10,10 & 0.053 & 0.052 & 0.051 &     & 10,10 & 0.050 & 0.051 & 0.049\\
      & 20,30 & 0.051 & 0.053 & 0.050 &     & 20,30 & 0.052 & 0.053 & 0.049\\
      & 30,30 & 0.056 & 0.050 & 0.049 &     & 30,30 & 0.049 & 0.047 & 0.048\\ \hline
\end{tabular}\caption{Estimated type I error for the test with critical region $V_k'>z_{1-\alpha}$. }
\label{tab:TRV2}
\end{table}

\subsection{The $U$-statistic to estimate $\|{\boldsymbol \pi}_1-{\boldsymbol \pi}_2\|^2$ } \label{u_est}
Let ${\boldsymbol N}_{1}=(n_{11}, \ldots, n_{1d})^\top  \sim   \mathcal{M}(n_{1.}; {\boldsymbol \pi}_{1})$,
${\boldsymbol N}_{2}=(n_{21}, \ldots, n_{2d})^\top  \sim   \mathcal{M}(n_{2.}; {\boldsymbol \pi}_{2})$, with
${\boldsymbol \pi}_{1}=(\pi_{11}, \ldots, \pi_{1d})^\top$, ${\boldsymbol \pi}_{2}=(\pi_{21}, \ldots, \pi_{2d})^\top \in \Xi_d$.
The $U$-statistic to estimate $\|{\boldsymbol \pi}_1-{\boldsymbol \pi}_2\|^2$ is
\[
T_{U_1}=\frac{1}{n_{1.}(n_{1.}-1)} \frac{1}{n_{2.}(n_{2.}-1)}\sum_{u\neq v} \sum_{s\neq t}h(X_{1u},X_{1v}; X_{2s}, X_{2t}),
\]
where $X_{11}, \ldots, X_{1n_{1.}}$ are iid so that $P(X_{11}=j)=\pi_{1j}$, $1\leq j \leq d$, and $X_{21}, \ldots, X_{2n_{2.}}$ are iid so that $P(X_{21}=j)=\pi_{2j}$, $1\leq j \leq d$, that is,  $X_{11}, \ldots, X_{1n_{1.}}$ are the data from population 1 (say) before summarizing them in the vector ${\boldsymbol N}_1$, and  $X_{21}, \ldots, X_{2n_{2.}}$ are the data from population 2 (say) before summarizing them in the vector ${\boldsymbol N}_2$; and $h$ is a two-sample symmetric kernel defined as
\begin{eqnarray*}
 h(X_{1u},X_{1v}; X_{2s}, X_{2t})  & =  & \sum_{j=1}^d \big\{I(X_{1u}=j)I(X_{1v}=j) + I(X_{2s}=j)I(X_{2t}=j) \\
&  &  -0.5 I(X_{1u}=j)I(X_{2s}=j)-0.5 I(X_{1u}=j)I(X_{2t}=j)\\ & &  -0.5 I(X_{1v}=j)I(X_{2s}=j)-0.5 I(X_{1v}=j)I(X_{2t}=j) \big\},
\end{eqnarray*}
where $I(\cdot )$  stands for the indicator function. From a practical point of view, $T_{U_1}$ can be expressed as in \eqref{TU1}.

\subsection{Asymptotic null distribution of the test statistic} \label{Asymptotic}
This subsection derives the asymptotic distribution of the test statistic $T_{U}$ under the null hypothesis. With this aim, it will be assumed that the sample sizes of the data from each population are comparable in the following sense:
\begin{equation}\label{sample.sizes}
n_{ir.}=c_{ir}m,  \quad 0<c_0\leq c_{ir}\leq C_0<\infty,  \quad i=1,2,\,\, r=1,\ldots, k,
\end{equation}
where $c_0$ and $C_0$ are two fixed constants. Notice that no assumption is made on $m$, it may remain bounded or increase arbitrarily. In the above expression, $c_{\blue{i}r}$ and $m$ are allowed to vary with $k$ so, strictly speaking, they should be denoted as  $c_{\blue{i}r}(k)$ and $m(k)$ but, in order to keep the notation as simple as possible, the dependence on $k$ is skipped.

\begin{theorem} \label{asymptotic.null.distribution}
Suppose that $n_{1r.}, \,n_{2r.} \geq 4 $, $\forall r$, that \eqref{sample.sizes} holds,  that
\begin{equation}\label{trazas}
{\rm tr}(\Sigma^2_{{\boldsymbol \pi}_{ir}})>\delta, \quad i=1,2,\,   \forall r,
\end{equation}
where $\delta$ is a positive constant, and that  $H_0$ is true.
Then
$T_{U}/\sqrt{var_0(T_{U})} \stackrel{\mathcal{L}}{\longrightarrow} Z$,   where $Z\sim N(0,1)$.
\end{theorem}

\begin{remark} \label{remark.trazas}
Let $ \boldsymbol \pi=(\pi_1, \ldots, \pi_d)^\top \in \Xi_d$, then
\begin{eqnarray}
{\rm tr}(\Sigma_{ \boldsymbol \pi}^2) & =    & \sum_{i=1}^d \pi_i^2-2\sum_{i=1}^d \pi_i^3+\left(\sum_{i=1}^d \pi_i^2\right)^2  \label{Sigma1} \\
                            & \geq & \sum_{i=1}^d \left\{\pi_i^2-2\pi_i^3+ \pi_i^4\right\} = \sum_{i=1}^d \{\pi_i(1-\pi_i)\}^2. \nonumber
                             %\geq  \{\pi_i(1-\pi_i)\}^2 \quad \forall i,
\end{eqnarray}
Thus,  \eqref{trazas} holds whenever $ \sum_{j=1}^d \{\pi_{irj}(1-\pi_{irj})\}^2>\zeta$, $i=1,2$,  $\forall r$, for some $\zeta>0$, which in turn is met if $\displaystyle \max_{1\leq j \leq d} \pi_{irj}\leq \epsilon$, $i=1,2$,  $\forall r$, for some $0<\epsilon<1$, which is not restrictive at all, it is just saying that there must be no population with a $\pi_{irj}$ equal to 1.
\end{remark}

\begin{remark} \label{remark.sample sizes}
Condition \eqref{sample.sizes} implies that all sample sizes are of the same order, all of them are either  small or large. Other conditions can be imposed on the sample sizes so that the convergence in Theorem \ref{asymptotic.null.distribution} holds true. An example is the following,
$$ \sum_{r=1}^k \left( \frac{1}{n_{1r.}(n_{1r.}-1)} + \frac{1}{n_{2r.}(n_{2r.}-1)} \right)
 \to \infty,$$
 which is fulfilled if   either $n_{1r.} \leq C k^a$ or  $n_{2r.} \leq C k^a$, $\forall r$, for some $0\leq a <1/2$ and some positive constant $C>0$, but no condition is imposed on how similar the sample sizes are.
\end{remark}

As commented in Section \ref{The.test.statistic}, for the result in Theorem \ref{asymptotic.null.distribution} to be useful to get a critical region for testing $H_0$, a consistent estimator of $var_0(T_{U})$ is needed. Next, we study the three estimators of $var_0(T_{U})$ introduced in that section.

\noindent \underline{Estimator 1.}

Let $\widehat{var}_0(T_{U})$ be as defined in Subsection \ref{test.1}. By construction,
\[E\left(\widehat{var}_0(T_{U})\right)= {var}_0(\chi^2_{U}),\]
under the null hypothesis and also under alternatives.
The next result shows that under certain weak assumptions, $\widehat{var}_0(T_{U})$  is a ratio consistent estimator of ${var}_0(T_{U})$.

\begin{theorem} \label{var0.consistencia}
Suppose that $n_{1r.}, \,n_{2r.} \geq 4 $, $\forall r$, that \eqref{sample.sizes} and \eqref{trazas} hold. Then
$\widehat{var}_0(T_{U})/ {var}_0(T_{U})  \stackrel{P}{\longrightarrow} 1$.
\end{theorem}

Observe that  Theorem \ref{var0.consistencia} is valid under the null hypothesis and under alternatives.

As an immediate consequence of  Theorems \ref{asymptotic.null.distribution} and \ref{var0.consistencia},  we have the following result.

\begin{corollary} \label{and0}
Suppose that assumptions in Theorem \ref{asymptotic.null.distribution}   hold. Then
$T_{U}/\sqrt{\widehat{var}_0(T_{U})} \stackrel{\mathcal{L}}{\longrightarrow} Z.$
\end{corollary}

\noindent \underline{Estimator 2.}
 Let $\widetilde{var}_0(T_{U})$ be as defined in  Subsection \ref{test.2}.
By construction,
  $E_0\left(\widetilde{var}_0(T_{U})\right)= {var}_0(T_{U})$.

The next result shows that, under the assumptions in Theorem  \ref{var0.consistencia} and assuming that the null hypothesis is true, $\widetilde{var}_0(T_{U})$  is also a ratio consistent estimator of ${var}_0(T_{U})$.

\begin{theorem} \label{var1.consistencia} Suppose that assumptions in Theorem \ref{var0.consistencia}   hold.  If $H_0$ is true then
$$\widetilde{var}_0(T_{U})/ {var}_0(T_{U})  \stackrel{P}{\longrightarrow} 1.$$
If $H_0$ is not true then $$\widetilde{var}_0(T_{U})/ {var}_{01}(T_{U})  \stackrel{P}{\longrightarrow} 1, $$
 where ${var}_{01}(T_{U})$ has the same expression as $var_{0}(T_{U})$ with $\Sigma_{ \boldsymbol \pi_{1r}}^2$ and $\Sigma_{ \boldsymbol \pi_{2r}}^2$  both replaced with $\Sigma_{ \boldsymbol \pi_{1r}}\Sigma_{ \boldsymbol \pi_{2r}}$, for each $r$.
\end{theorem}

As an immediate consequence of  Theorems \ref{asymptotic.null.distribution} and \ref{var1.consistencia},   we have the following result.

\begin{corollary} \label{and1}
Suppose that assumptions in Theorem \ref{asymptotic.null.distribution}   hold. Then
$T_{U}/\sqrt{\widetilde{var}_0(T_{U})} \stackrel{\mathcal{L}}{\longrightarrow} Z.$
\end{corollary}

\noindent \underline{Estimator 3.}
 Let $\overline{var}_0(T_{U})$ be as defined in Subsection \ref{test.3}. By construction,
$E_0\left(\overline{var}_0(T_{U})\right)= {var}_0(T_{U}).$

As with the previous estimators, under the assumptions in Theorem  \ref{var0.consistencia} and assuming that the null hypothesis is true, $\overline{var}_0(T_{U})$  is also a ratio consistent estimator of ${var}_0(T_{U})$.

\begin{theorem} \label{var2.consistencia}
Suppose that assumptions in Theorem \ref{var0.consistencia}   hold.  If $H_0$ is true then
$$\overline{var}_0(T_{U})/ {var}_0(T_{U})  \stackrel{P}{\longrightarrow} 1.$$
If $H_0$ is not true, and $n_{1r.}/n_{.r.} \to \lambda_{r} \in (0,1)$, for each $r$, then
 $$\overline{var}_0(T_{U})/ {var}_{02}(T_{U})  \stackrel{P}{\longrightarrow} 1,$$
 where ${var}_{02}(T_{U})$ has the same expression as ${var}_{0}(T_{U})$ with $\Sigma_{ \boldsymbol \pi_{1r}}$ and $\Sigma_{ \boldsymbol \pi_{2r}}$ replaced with
 $\Sigma_{ \boldsymbol \pi_{r}}$, where $ \boldsymbol \pi_{r}=\lambda_{r}\Sigma_{\pi_{1r}}+(1- \lambda_{r})\Sigma_{ \boldsymbol \pi_{2r}}$, for each $r$.
\end{theorem}

From \eqref{sample.sizes}, it follows that $0<c_0/(c_0+C_0) \leq \lambda_{r} \leq C_0/(c_0+C_0)<1$, for each $r$.

As an immediate consequence of  Theorems \ref{asymptotic.null.distribution} and \ref{var2.consistencia}, we have the following result.

\begin{corollary} \label{and2}
Suppose that assumptions in Theorem \ref{asymptotic.null.distribution}   hold. Then
$T_{U}/\sqrt{\overline{var}_0(T_{U})} \stackrel{\mathcal{L}}{\longrightarrow} Z.$
\end{corollary}

\subsection{Asymptotic power}\label{a_power}
 This section studies the asymptotic power of the proposed tests. To this end, we first derive the asymptotic distribution of the test statistic $T_{U}$ under alternatives. With this aim, w.l.o.g. it will be assumed that
 \begin{equation} \label{alternatives}
 {\boldsymbol \pi}_{1r} \neq {\boldsymbol \pi}_{1r}, \,\, 1\leq r \leq v, \quad {\boldsymbol \pi}_{1r} = {\boldsymbol \pi}_{1r}, \,\, v+1\leq r \leq k,
 \end{equation}
for some $1\leq v \leq k$. Here, $v$ is allowed to depend on $k$, $v=v(k)$,  but to simplify notation, such dependence is skipped.

\begin{theorem} \label{asymptotic.distribution}
Suppose that $n_{1r.}, \,n_{2r.} \geq 4 $, $\forall r$, that \eqref{sample.sizes}, \eqref{alternatives} hold,  that
\begin{equation}\label{non.ort}
\max \left\{ ({\boldsymbol \pi}_{1r}-{\boldsymbol \pi}_{2r})^\top \Sigma_{{\boldsymbol \pi}_{1r}}({\boldsymbol \pi}_{1r}-{\boldsymbol \pi}_{2r}),\,
({\boldsymbol \pi}_{1r}-{\boldsymbol \pi}_{2r})^\top \Sigma_{{\boldsymbol \pi}_{2r}}({\boldsymbol \pi}_{1r}-{\boldsymbol \pi}_{2r}) \right\} >\delta, \quad 1 \leq r \leq v,
 \end{equation}
where $\delta$ is a positive constant, and that  $v/k\to p\in (0,1]$.
Then
$\{T_{U}-E(T_{U})\}/\sqrt{var(T_{U})} \stackrel{\mathcal{L}}{\longrightarrow} Z$.
\end{theorem}

Let ${\boldsymbol \pi}_1=(\pi_{11}, \ldots, \pi_{1d})^\top, \, {\boldsymbol \pi}_2=(\pi_{21}, \ldots, \pi_{2d})^\top \in \Xi_d$, and assume that $({\boldsymbol \pi}_{1}-{\boldsymbol \pi}_{2})^\top \Sigma_{{\boldsymbol \pi}_{1}}({\boldsymbol \pi}_{1}-{\boldsymbol \pi}_{2})>\delta$, for some $\delta>0$, that is,
\[ \sum_{i=1}^d(\pi_{1i}-\pi_{2i})^2\pi_{1i}-\left\{ \sum_{i=1}^d(\pi_{1i}-\pi_{2i})\pi_{1i}\right\}^2>\delta. \]
It entails that $\|{\boldsymbol \pi}_1-{\boldsymbol \pi}_2\|^2>\delta$, since
\begin{equation} \label{cota.inf}
\|{\boldsymbol \pi}_1-{\boldsymbol \pi}_2\|^2 \geq  \sum_{i=1}^d(\pi_{1i}-\pi_{2i})^2\pi_{1i}>\delta+\left\{ \sum_{i=1}^d(\pi_{1i}-\pi_{2i})\pi_{1i}\right\}^2 \geq \delta.
\end{equation}

 \noindent \underline {Power of Test 1} Let us consider
 the test that rejects $H_0$ when
$T_{U}/\sqrt{\widehat{var}_0(T_{U})}  \geq z_{1-\alpha},$
for some $\alpha \in (0,1)$.
Suppose that  the assumptions in Theorems   \ref{var0.consistencia} and  \ref{asymptotic.distribution} hold, then
 \begin{equation} \label{aux.pot1}
 P\left(\frac{T_{U}}{\sqrt{\widehat{var}_0(T_{U})} } >z_{1-\alpha} \right) \approx \Phi \left( \frac{\sqrt{m}E(T_{U})}{\sqrt{{var}(\sqrt{m}T_{U})}} +\frac{\sqrt{{var}_0(T_{U})}}{\sqrt{{var}(T_{U})}} z_{\alpha}\right).
 \end{equation}
 First, notice that
\begin{equation} \label{facil}
{var}_0(T_{U})/{var}(T_{U})\leq 1.
\end{equation}
On the other hand,   from \eqref{aux2.2} and  \eqref{aux2.3}  in the proof of Theorem \ref{asymptotic.distribution}, we have that
\begin{equation}\label{aux2.1}
{var}(\sqrt{m}T_{U})=\sum_{j=1}^k var(\sqrt{m}T_{U_r}) \leq Ck
\end{equation}
where $C$ is a positive constant (depending on $\delta$, $c_0$ and $C_0$).
From  \eqref{non.ort} and \eqref{cota.inf}, we also have that
\begin{equation}\label{medias}
\sqrt{m}E(T_{U}) \geq \sqrt{m}v\delta.
\end{equation}
Finally, from \eqref{facil}-\eqref{medias},
\begin{equation}\label{cota1}
\frac{\sqrt{m}E(T_{U})}{\sqrt{{var}(\sqrt{m}T_{U})}} +\frac{\sqrt{{var}_0(T_{U})}}{\sqrt{{var}(T_{U})}} z_{\alpha}
\geq C \sqrt{m} \sqrt{v/k}\sqrt{v}-|z_{\alpha}|,
\end{equation}
where $C$ is a positive constant (which can differ from that in \eqref{aux2.1}). Since  $v/k\to p\in (0,1]$, which entails $v\to \infty$, from
 \eqref{aux.pot1} and \eqref{cota1}, it readily follows that the power of the test goes to 1. This result is formally stated in the  following theorem.

\begin{theorem} \label{power.test1}
Let $\alpha \in (0,1)$. If assumptions in Theorems   \ref{var0.consistencia} and  \ref{asymptotic.distribution} hold, then
$$P\left(\frac{T_{U}}{\sqrt{\widehat{var}_0(T_{U})} } >z_{1-\alpha} \right) \to 1.$$
\end{theorem}

 \noindent \underline {Power of Test 2} Let us consider
 the test that rejects $H_0$ when
$T_{U}/\sqrt{\widetilde{var}_0(T_{U})}  \geq z_{1-\alpha},$
for some $\alpha \in (0,1)$.
Suppose that  the assumptions in  Theorems   \ref{var1.consistencia} and  \ref{asymptotic.distribution}  hold, then
\[ 
 P\left(\frac{T_{U}}{\sqrt{\widetilde{var}_0(T_{U})} } >z_{1-\alpha} \right) \approx \Phi \left( \frac{E(T_{U})}{\sqrt{{var}(T_{U})}} +\frac{\sqrt{{var}_{01}(T_{U})}}{\sqrt{{var}(T_{U})}} z_{\alpha}\right).
\] 

To derive the power, it is needed to study the quotient ${var}_{01}(T_{U})/{var}(T_{U})$.

 \begin{lemma} \label{coc.var2}
 Assume  that there exists a  constant $M (\geq 1)$ such that
  $\max \{ {\rm tr}(\Sigma_{\pi_{1r}}^2), {\rm tr}(\Sigma_{\pi_{2r}}^2) \} \leq M
  \min \{ {\rm tr}(\Sigma_{\pi_{1r}}^2), {\rm tr}(\Sigma_{\pi_{2r}}^2) \}$, for all $r$. Then
 $0 \leq {var}_{01}(T_{U})/{var}(T_{U}) \leq {var}_{01}(T_{U})/{var_0}(T_{U})  \leq M. $
\end{lemma}

Proceeding as before (for Test 1), if assumptions Theorems   \ref{var1.consistencia} and  \ref{asymptotic.distribution}   and Lemma
\ref{coc.var2} hold, then it is obtained that
\[ 
\frac{E(T_{U})}{\sqrt{{var}(T_{U})}} +\frac{\sqrt{{var}_{01}(T_{U})}}{\sqrt{{var}(T_{U})}} z_{\alpha} \geq
C \sqrt{m} \sqrt{v/k}\sqrt{v}-M|z_{\alpha}|,
\] 
where $C$ is a positive constant and $M$ is the constant in Lemma \ref{coc.var2}. The right-hand side of the above inequality goes to $\infty$, and hence, under the stated assumptions,  the test is consistent.  This result is formally expressed in the  following theorem.

\begin{theorem} \label{power.test2}
Let $\alpha \in (0,1)$. If assumptions in Theorems   \ref{var1.consistencia} and  \ref{asymptotic.distribution} and Lemma \ref{coc.var2}  hold, then
$$P\left(\frac{T_{U}}{\sqrt{\widetilde{var}_0(T_{U})} } >z_{1-\alpha} \right) \to 1.$$
\end{theorem}

\noindent \underline {Power of Test 3} Let us consider
 the test that rejects $H_0$ when
$T_{U}/\sqrt{\overline{var}_0(T_{U})}  \geq z_{1-\alpha},$
for some $\alpha \in (0,1)$.
Suppose that  the assumptions in Theorems   \ref{var2.consistencia} and \ref{asymptotic.distribution}  hold, then
 \[ %\begin{equation} \label{aux.pot3}
 P\left(\frac{T_{U}}{\sqrt{\overline{var}_0(T_{U})} } >z_{1-\alpha} \right) \approx \Phi \left( \frac{E(T_{U})}{\sqrt{{var}(T_{U})}} +\frac{\sqrt{{var}_{02}(T_{U})}}{\sqrt{{var}(T_{U})}} z_{\alpha}\right).
 \] %\end{equation}

To obtain the power, the key point is the study of the quantity  ${var}_{02}(T_{U})/{var}(T_{U})$.

 \begin{lemma} \label{coc.var3}
 Assume that the assumptions in Lemma \ref{coc.var2} hold.
Then
  $0 \leq {var}_{02}(T_{U})/{var}(T_{U}) \leq $ $ {var}_{02}(T_{U})/{var_0}(T_{U}) \leq M+1. $
 \end{lemma}

In consequence, proceeding as before (for Test 2), if assumptions Theorems   \ref{var2.consistencia} and  \ref{asymptotic.distribution}   and Lemma
\ref{coc.var3} hold, then the test is consistent.  This result is formally stated in the  following theorem.

\begin{theorem} \label{power.test3}
Let $\alpha \in (0,1)$. If assumptions in Theorems   \ref{var2.consistencia} and  \ref{asymptotic.distribution} and Lemma \ref{coc.var3}  hold, then
$$P\left(\frac{T_{U}}{\sqrt{\overline{var}_0(T_{U})} } >z_{1-\alpha} \right) \to 1.$$
\end{theorem}

\subsection{Proofs} \label{proofs}
Along this section $C$ is a generic positive constant taking many different values
throughout the proofs.

\noindent {\bf Proof of Theorem \ref{asymptotic.null.distribution}} \hspace{2pt}
Under the null hypothesis ${\boldsymbol \pi}_{1r}={\boldsymbol \pi}_{2r}:={\boldsymbol \pi}_{r}$, $\forall r$, and thus
\begin{equation}\label{aux1.5}
var_0(mT_{U_r})={\rm tr} \left( \Sigma_{ {\boldsymbol \pi}_{r} }^2 \right)\left\{\frac{2m^2}{n_{1r.}(n_{1r.}-1)} + \frac{2m^2}{n_{2r.}(n_{2r.}-1)} + \frac{4m^2}{n_{1r.}n_{2r.}} \right\}.
\end{equation}
Let ${\boldsymbol \pi}=(\pi_1, \ldots, \pi_d)^\top \in \Xi_d$, from \eqref{Sigma1} it follows that
\begin{equation}\label{aux1.6}
{\rm tr} \left( \Sigma_{ {\boldsymbol \pi} }^2 \right) \leq
\sum_{i=1}^d \pi_i^2+\left(\sum_{i=1}^d \pi_i^2\right)^2 \leq \sum_{i=1}^d \pi_i+\left(\sum_{i=1}^d \pi_i\right)^2 =2.
\end{equation}
From \eqref{sample.sizes}, \eqref{trazas},  \eqref{aux1.5},  \eqref{aux1.6} and taking into account that if $n_{ir.} \geq 4$, then $1 \leq  n_{ir.}/(n_{ir.}-1) \leq 4/3$, we get that
\begin{equation}\label{aux1.1}
\frac {8\delta}{C_0^2} \leq var_0(mT_{U_r})=E_0(m^2T_{U_r}^2) \leq  \frac{8\times4 \times 2}{3c_0^2}, \quad \forall r.
\end{equation}

Notice that

$$\frac{T_{U}}{ \sqrt{var_0(T_{U})}}=\frac {\sum_{r=1}^k m T_{U_r} }{ \left \{ \sum_{r=1}^k var_0(m T_{U_r}) \right \}^{1/2} }.
$$
To show the result, we will see  that the Lindeberg condition
\[
\ell_k(\varepsilon)=\frac{ \displaystyle \sum_{r=1}^kE_0 \left[m^2T_{U_r}^2I\left\{ m^2T_{U_r}^2>\varepsilon \sum_{r=1}^k var (mT_{U_r})\right\}\right] }{ \displaystyle \sum_{r=1}^k var_0 (mT_{U_r}) } \to 0,\quad \forall \varepsilon>0,
\]
is met. Then, the result follows from Theorem 1.9.2.A in Serfling \cite{Serfling}.

From \eqref{aux1.1},
\begin{equation}\label{aux1.2}
\sum_{r=1}^k var (mT_{U_r}) \geq Ck,
\end{equation}
and hence,
\begin{equation}\label{aux1.3}
E_0 \left[  m^2T_{U_r}^2I \left\{ m^2T_{U_r}^2>\varepsilon \sum_{r=1}^k var (mT_{U_r}) \right\} \right] \leq
E_0\left[ m^2T_{U_r}^2I\left\{ m^2T_{U_r}^2>\varepsilon Ck   \right\}\right].
\end{equation}
From \eqref{aux1.1},
 $E_0(m^2T_{U_r}^2)<\infty$, $\forall r$,  and hence
 \begin{equation}\label{aux1.4}
E_0\left[m^2T_{U_r}^2I\left\{ m^2T_{U_r}^2>\varepsilon Ck \right\}\right] \to 0,\quad  \forall \varepsilon>0,  \,\,\, \forall r.
\end{equation}
Finally, from \eqref{aux1.2}--\eqref{aux1.4}, it is concluded that $\ell_k(\varepsilon) \to 0$,  $\forall \varepsilon>0.$
$\Box$

\medskip
\noindent {\bf Proof of Theorem \ref{var0.consistencia}} \hspace{2pt} Let
 \begin{eqnarray*}
\hat{\sigma}_{0r}^2 & =  & \frac{2}{n_{1r.}(n_{1r.}-1)} {\rm tr}(\widehat{\Sigma^2}_{{\boldsymbol \pi}_{1r}}) + \frac{2}{n_{2r.}(n_{2r.}-1)} {\rm tr}(\widehat{\Sigma^2}_{{\boldsymbol \pi}_{2r}}) +
  \frac{4}{n_{1r.}n_{2r.}} {\rm tr}(\widehat{\Sigma}_{{\boldsymbol \pi}_{1r}}
 \widehat{\Sigma}_{{\boldsymbol \pi}_{2r}}).
 \end{eqnarray*}
With this notation, $\widehat{var}_0(T_{U})=\frac{1}{k} \sum_{r=1}^k \hat{\sigma}_{0r}^2$.
Let $\epsilon >0$, by Markov inequality,
\[
P\left( \left| \widehat{var}_0(T_{U})-{var}_0(T_{U}) \right|>\epsilon {var}_0(T_{U}) \right) \leq \frac{1}{\epsilon^2 {var}_0^2(T_{U})} E \left\{ \left( \widehat{var}_0(T_{U})- {var}_0(T_{U}) \right)^2 \right\}.
\]
By construction, $E\left\{ \hat{\sigma}_{0r}^2 \right\}={var}_0(T_{U_r})$, and thus
\[
 E \left\{ \left( \widehat{var}_0(T_{U})- {var}_0(T_{U}) \right)^2 \right\}=\frac{1}{k^2} \sum_{r=1}^k var( \hat{\sigma}_{0r}^2 ).
 \]
Tedious, but routine calculations, show that
\[ 
var \left( {\rm tr}(\widehat{\Sigma^2}_{{\boldsymbol \pi}_{1r}}) \right) \leq C  \frac{1}{n_{1r.}}, \quad
var \left( {\rm tr}(\widehat{\Sigma^2}_{{\boldsymbol \pi}_{2r}}) \right) \leq C  \frac{1}{n_{2r.}}, \quad
var \left( (\widehat{\Sigma}_{{\boldsymbol \pi}_{1r}} \widehat{\Sigma}_{{\boldsymbol \pi}_{2r}}) \right) \leq C  \left(\frac{1}{n_{1r.}}+ \frac{1}{n_{2r.}}\right),
\] 
and thus
\[
var\left(\hat{\sigma}_{0r}^2 \right) \leq C  \frac{1}{m^5}. %\frac{1}{n_{1r.}^5 \wedge n_{2r.}^5}.
\]
On the other hand,
$${var}_0(T_{U}) \geq C \frac{1}{m^2}. %\left(\frac{1}{n_{1r.}^2}+\frac{1}{n_{2r.}^2}\right)  \geq C \frac{1}{n_{1r.}^2\vee n_{2r.}^2}.
$$
Therefore,
\[
\frac{E \left\{ \left( \widehat{var}_0(T_{U})- {var}_0(T_{U}) \right)^2 \right\}}{ {var}_0^2(T_{U})} \leq
C \frac{1}{km},
\]
which converges to 0, implying that  $\widehat{var}_0(T_{U})/{var}_0(T_{U}) \stackrel{P}{\longrightarrow} 1$. $\Box$

%-----------------------

\medskip
\noindent {\bf Proof of Theorems \ref{var1.consistencia} and  \ref{var2.consistencia} } \hspace{2pt} Their proofs have similar steps to those given in the proof of
 Theorem \ref{var0.consistencia}, so we omit them.  $\Box$

\medskip

%-----------------------

\noindent {\bf Proof of Theorem \ref{asymptotic.distribution}} \hspace{2pt}
From \eqref{sample.sizes} and \eqref{non.ort}, it follows that
\begin{equation}\label{aux22.1}
{var}(\sqrt{m}T_{U})=\sum_{j=1}^k var(\sqrt{m}T_{U_r}) \geq \sum_{j=1}^v var(\sqrt{m}T_{U_r}) \geq v\frac{4\delta}{C_0}.
\end{equation}
On the other hand,
using \eqref{sample.sizes},  \eqref{aux1.6}, and that for
${\boldsymbol \pi}_1=(\pi_{11}, \ldots, \pi_{1d})^\top, \, {\boldsymbol \pi}_2=(\pi_{21}, \ldots, \pi_{2d})^\top  \in \Xi_d$,
\[
({\boldsymbol \pi}_1-{\boldsymbol \pi}_2)^\top \Sigma_{ {\boldsymbol \pi}_1} ({\boldsymbol \pi}_1-{\boldsymbol \pi}_2) \leq \sum_{i=1}^d\pi_{1j}
(\pi_{1j}-\pi_{2j})^2\leq 2.
\]
and
\[
{\rm tr}(\Sigma_{ {\boldsymbol \pi}_1} \Sigma_{ {\boldsymbol \pi}_1}) \leq {\rm tr}^{1/2}(\Sigma_{ {\boldsymbol \pi}_1}^2)
{\rm tr}^{1/2}(\Sigma_{ {\boldsymbol \pi}_2}^2) \leq 2,
\]
we get that
\begin{equation}\label{aux2.2}
var( \sqrt{m}T_{U_r}) \leq \frac{C}{c_0}, \quad 1\leq r \leq v,
\end{equation}
and
\begin{equation}\label{aux2.3}
var( \sqrt{m}T_{U_r}) \leq \frac{C}{c_0}, \quad v+1\leq r \leq k.
\end{equation}
We have  that
$$
\frac{T_{U}-E(T_{U})}{ \sqrt{var(T_{U})}}=\frac {\sum_{r=1}^k \sqrt{m} \left\{ T_{U_r}-E(T_{U_r})\right\} }{ \left \{ \sum_{r=1}^k var( \sqrt{m}T_{U_r}) \right \}^{1/2} }.
$$
To show the result, we will see  that the Lindeberg condition
\[
\ell_k(\varepsilon)=\frac{ \displaystyle \sum_{r=1}^kE \left[m \left\{T_{U_r}-E(T_{U_r})\right\}^2I\left\{ m \left\{T_{U_r}-E(T_{U_r})\right\}^2>\varepsilon \sum_{r=1}^k var (\sqrt{m}T_{U_r})\right\}\right] }{ \displaystyle \sum_{r=1}^k var (\sqrt{m}T_{U_r}) } \to 0,\quad \forall \varepsilon>0,
\]
is met. From \eqref{aux22.1}-\eqref{aux2.3}, and noticing that  $v/k\to p\in(0,1]$ entails that $v\to \infty$, we have that
\begin{equation}\label{aux2.4}
\begin{array}{c}
\displaystyle
E \left[m \left\{T_{U_r}-E(T_{U_r})\right\}^2I\left\{ m \left\{T_{U_r}-E(T_{U_r})\right\}^2>\varepsilon \sum_{r=1}^k var (\sqrt{m}T_{U_r})\right\}\right]\leq \\
\displaystyle
E \left[m \left\{T_{U_r}-E(T_{U_r})\right\}^2I\left\{ m \left\{T_{U_r}-E(T_{U_r})\right\}^2>\varepsilon
 v 4\delta/C_0
\right\}\right] \to 0, \quad  \forall \varepsilon>0,  \,\,\, \forall r.
\end{array}
\end{equation}
Finally, from \eqref{aux22.1} and  \eqref{aux2.4}, it is concluded that $\ell_k(\varepsilon) \to 0$,  $\forall \varepsilon>0.$
$\Box$

\medskip
\noindent {\bf Proof of Lemma \ref{coc.var2}} \hspace{2pt}  Assume that
${\rm tr}(\Sigma^2_{{\boldsymbol \pi}_{1r}}) \geq {\rm tr}(\Sigma^2_{{\boldsymbol \pi}_{2r}})$, then
\begin{equation}\label{las.Ms}
{\rm tr}(\Sigma_{{\boldsymbol \pi}_{1r}} \Sigma_{{\boldsymbol \pi}_{2r}}) \leq {\rm tr}^{1/2}(\Sigma^2_{{\boldsymbol \pi}_{1r}})
 {\rm tr}^{1/2}(\Sigma^2_{{\boldsymbol \pi}_{2r}}) \leq \left \{ \begin{array}{l}
 {\rm tr}(\Sigma^2_{{\boldsymbol \pi}_{1r}})  \leq M  {\rm tr}(\Sigma^2_{{\boldsymbol \pi}_{1r}}), %\quad \mbox{because $M \geq 1$,}
 \\
 M {\rm tr}(\Sigma^2_{{\boldsymbol \pi}_{2r}}).
\end{array}\right.
\end{equation}
Therefore,
\begin{eqnarray*}
var_{01}(T_U) & = & \frac{1}{k}\sum_{r=1}^k{\rm tr}(\Sigma_{{\boldsymbol \pi}_{1r}} \Sigma_{{\boldsymbol \pi}_{2r}})\left(
\frac{2}{n_{1r.}(n_{1r.}-1)}  + \frac{2}{n_{2r.}(n_{2r.}-1)} +
  \frac{4}{n_{1r.}n_{2r.}}
\right) \\ & \leq &  M {var}_0(T_{U}) \leq M {var}(T_{U}),
\end{eqnarray*}
and hence, $0 \leq  \frac{{var}_{01}(T_{U})}{{var_0}(T_{U})} \leq \frac{{var}_{01}(T_{U})}{{var}(T_{U})} \leq M. $ $\Box$

\medskip

\noindent {\bf Proof of Lemma \ref{coc.var3}} \hspace{2pt}  Let
$\lambda_r=n_{1r.}/n_{.r.}$, $1 \leq r \leq k$. Assume that
${\rm tr}(\Sigma^2_{{\boldsymbol \pi}_{1r}}) \geq {\rm tr}(\Sigma^2_{{\boldsymbol \pi}_{2r}})$, then from  \eqref{las.Ms},
$${\rm tr}\left[ \left\{ \lambda_r \Sigma_{{\boldsymbol \pi}_{1r}}+(1-\lambda_r)\Sigma_{{\boldsymbol \pi}_{2r}} \right\}^2 \right] \leq M{\rm tr}(\Sigma_{{\boldsymbol \pi}_{2r}}^2).$$
Moreover, since
$$
\frac{1}{ n_{1r.} n_{2r.} }\leq \frac{1}{ n_{1r.} (n_{1r.}-1) }+
\frac{1}{n_{2r.}(n_{2r.}-1)}
$$
and ${\rm tr}(\Sigma^2_{{\boldsymbol \pi}_{1r}}\Sigma^2_{{\boldsymbol \pi}_{2r}}) \geq 0$, one gets that
\begin{eqnarray*}
var_{02}(T_U) & = & \frac{1}{k}\sum_{r=1}^k\left(
\frac{2}{n_{1r.}(n_{1r.}-1)}+ \frac{2}{n_{2r.}(n_{2r.}-1)}+
\frac{4}{n_{1r.}n_{2r.}} \right)
{\rm tr}\left[ \left\{ \lambda_r \Sigma_{{\boldsymbol \pi}_{1r}}+(1-\lambda_r)\Sigma_{{\boldsymbol \pi}_{2r}} \right\}^2 \right]\\
& \leq & (M+1) \frac{1}{k}\sum_{r=1}^k\left(
\frac{2}{n_{1r.}(n_{1r.}-1)}+ \frac{2}{n_{2r.}(n_{2r.}-1)} \right)
{\rm tr}(\Sigma^2_{{\boldsymbol \pi}_{2r}})\\
& \leq & (M+1)var_0(T_U) \leq  (M+1)var(T_U).
\end{eqnarray*}
and hence, $0 \leq  \frac{{var}_{02}(T_{U})}{{var_0}(T_{U})} \leq \frac{{var}_{02}(T_{U})}{{var}(T_{U})} \leq M+1.$ $\Box$

\subsection{Simulation results for the level} \label{level}
\begin{table}[h] %[p]
\centering
\begin{tabular}{ccccc|ccccc}
  $k$ & $n_{1r},n_{2r}$ & Test 1 & Test 2 & Test 3 & $k$ & $n_{1r},n_{2r}$ & Test 1 & Test 2 & Test 3\\ \hline
  20  & 5,10  & 0.054 & 0.066 & 0.055 & 200 & 5,10  & 0.054 & 0.055 & 0.055 \\
      & 10,10 & 0.057 & 0.062 & 0.055 &     & 10,10 & 0.054 & 0.054 & 0.053 \\
      & 20,30 & 0.054 & 0.056 & 0.054 &     & 20,30 & 0.050 & 0.051 & 0.050 \\
      & 30,30 & 0.054 & 0.055 & 0.054 &     & 30,30 & 0.051 & 0.052 & 0.051 \\
  50  & 5,10  & 0.053 & 0.058 & 0.054 & 500 & 5,10  & 0.053 & 0.055 & 0.054 \\
      & 10,10 & 0.052 & 0.057 & 0.052 &     & 10,10 & 0.053 & 0.055 & 0.053 \\
      & 20,30 & 0.052 & 0.054 & 0.052 &     & 20,30 & 0.054 & 0.054 & 0.054 \\
      & 30,30 & 0.054 & 0.055 & 0.054 &     & 30,30 & 0.054 & 0.054 & 0.054 \\
  100 & 5,10  & 0.051 & 0.057 & 0.052 & 750 & 5,10  & 0.055 & 0.053 & 0.055 \\
      & 10,10 & 0.051 & 0.055 & 0.051 &     & 10,10 & 0.051 & 0.052 & 0.051 \\
      & 20,30 & 0.052 & 0.053 & 0.052 &     & 20,30 & 0.050 & 0.051 & 0.050 \\
      & 30,30 & 0.054 & 0.055 & 0.054 &     & 30,30 & 0.050 & 0.050 & 0.050 \\
  \hline
\end{tabular}\caption{Estimated type I error for the Setting 1 ($d=10$).} \label{tab:3}
\end{table}

\begin{table}
\centering
\begin{tabular}{ccccc|ccccc}
  $k$ & $n_{1r},n_{2r}$ & Test 1 & Test 2 & Test 3& $k$ & $n_{1r},n_{2r}$ & Test 1 & Test 2 & Test 3\\ \hline
  20  & 5,10  & 0.047 & 0.069 & 0.054 & 200 & 5,10  & 0.049 & 0.056 & 0.052 \\
      & 10,10 & 0.051 & 0.062 & 0.050 &     & 10,10 & 0.051 & 0.054 & 0.051 \\
      & 20,30 & 0.055 & 0.060 & 0.056 &     & 20,30 & 0.051 & 0.052 & 0.051 \\
      & 30,30 & 0.051 & 0.054 & 0.051 &     & 30,30 & 0.050 & 0.051 & 0.050 \\
  50  & 5,10  & 0.050 & 0.062 & 0.053 & 500 & 5,10  & 0.048 & 0.056 & 0.052 \\
      & 10,10 & 0.055 & 0.063 & 0.055 &     & 10,10 & 0.051 & 0.053 & 0.051 \\
      & 20,30 & 0.051 & 0.054 & 0.051 &     & 20,30 & 0.051 & 0.052 & 0.051 \\
      & 30,30 & 0.051 & 0.053 & 0.051 &     & 30,30 & 0.045 & 0.045 & 0.045 \\
  100 & 5,10  & 0.046 & 0.054 & 0.049 & 750 & 5,10  & 0.052 & 0.055 & 0.054 \\
      & 10,10 & 0.054 & 0.057 & 0.052 &     & 10,10 & 0.052 & 0.053 & 0.052 \\
      & 20,30 & 0.050 & 0.052 & 0.050 &     & 20,30 & 0.053 & 0.054 & 0.054 \\
      & 30,30 & 0.052 & 0.053 & 0.052 &     & 30,30 & 0.048 & 0.048 & 0.047 \\
  \hline
\end{tabular}\caption{Estimated type I error for the Setting 1 ($d=20$).} \label{tab:4}
\end{table}

\begin{table}
\centering
\begin{tabular}{ccccc|ccccc}
  $k$ & $n_{1r},n_{2r}$ & Test 1 & Test 2 & Test 3& $k$ & $n_{1r},n_{2r}$ & Test 1 & Test 2 & Test 3\\ \hline
  20  & 5,10  & 0.067 & 0.072 & 0.062 & 200 & 5,10  & 0.058 & 0.060 & 0.058 \\
      & 10,10 & 0.056 & 0.060 & 0.053 &     & 10,10 & 0.057 & 0.058 & 0.055 \\
      & 20,30 & 0.058 & 0.060 & 0.058 &     & 20,30 & 0.056 & 0.056 & 0.055 \\
      & 30,30 & 0.064 & 0.064 & 0.062 &     & 30,30 & 0.052 & 0.052 & 0.052 \\
  50  & 5,10  & 0.061 & 0.065 & 0.058 & 500 & 5,10  & 0.051 & 0.052 & 0.051 \\
      & 10,10 & 0.058 & 0.060 & 0.057 &     & 10,10 & 0.052 & 0.053 & 0.052 \\
      & 20,30 & 0.059 & 0.060 & 0.058 &     & 20,30 & 0.056 & 0.056 & 0.056 \\
      & 30,30 & 0.055 & 0.056 & 0.055 &     & 30,30 & 0.048 & 0.048 & 0.048 \\
  100 & 5,10  & 0.053 & 0.055 & 0.051 & 750 & 5,10  & 0.051 & 0.052 & 0.054 \\
      & 10,10 & 0.056 & 0.058 & 0.055 &     & 10,10 & 0.055 & 0.056 & 0.055 \\
      & 20,30 & 0.056 & 0.056 & 0.056 &     & 20,30 & 0.053 & 0.053 & 0.053 \\
      & 30,30 & 0.053 & 0.056 & 0.055 &     & 30,30 & 0.050 & 0.049 & 0.050 \\
  \hline
\end{tabular}\caption{Estimated type I error for the Setting 2 ($d=5$).} \label{tab:5}
\end{table}

\begin{table}
\centering
\begin{tabular}{ccccc|ccccc}
  $k$ & $n_{1r},n_{2r}$ & Test 1 & Test 2 & Test 3& $k$ & $n_{1r},n_{2r}$ & Test 1 & Test 2 & Test 3\\ \hline
  20  & 5,10  & 0.062 & 0.069 & 0.062 & 200 & 5,10  & 0.054 & 0.056 & 0.053 \\
      & 10,10 & 0.062 & 0.066 & 0.059 &     & 10,10 & 0.050 & 0.052 & 0.050 \\
      & 20,30 & 0.059 & 0.060 & 0.058 &     & 20,30 & 0.051 & 0.052 & 0.051 \\
      & 30,30 & 0.064 & 0.065 & 0.062 &     & 30,30 & 0.052 & 0.052 & 0.052 \\
  50  & 5,10  & 0.056 & 0.060 & 0.053 & 500 & 5,10  & 0.055 & 0.056 & 0.055 \\
      & 10,10 & 0.056 & 0.058 & 0.055 &     & 10,10 & 0.050 & 0.051 & 0.050 \\
      & 20,30 & 0.058 & 0.059 & 0.057 &     & 20,30 & 0.053 & 0.053 & 0.053 \\
      & 30,30 & 0.056 & 0.057 & 0.056 &     & 30,30 & 0.053 & 0.053 & 0.053 \\
  100 & 5,10  & 0.052 & 0.055 & 0.051 & 750 & 5,10  & 0.055 & 0.056 & 0.055 \\
      & 10,10 & 0.055 & 0.056 & 0.054 &     & 10,10 & 0.053 & 0.054 & 0.053 \\
      & 20,30 & 0.052 & 0.053 & 0.052 &     & 20,30 & 0.051 & 0.052 & 0.052 \\
      & 30,30 & 0.054 & 0.055 & 0.054 &     & 30,30 & 0.054 & 0.054 & 0.054 \\
  \hline
\end{tabular}\caption{Estimated type I error for the Setting 2 ($d=10$).} \label{tab:6}
\end{table}

\begin{table}
\centering

\begin{tabular}{cccccc|cccccc}
  $k$ & $n_{1r},n_{2r}$ &  Test 4 & Test 5 & Test 6 & Test 7 & $k$ & $n_{1r},n_{2r}$ & Test 4 & Test 5 & Test 6 & Test 7\\ \hline
  20  & 5,10  & 0.065 & 0.091 & 0.055 & 0.055 & 200 & 5,10  & 0.055 & 0.086 & 0.051 & 0.051 \\
      & 10,10 & 0.064 & 0.081 & 0.056 & 0.055 &     & 10,10 & 0.055 & 0.077 & 0.049 & 0.049 \\
      & 20,30 & 0.062 & 0.068 & 0.056 & 0.056 &     & 20,30 & 0.054 & 0.058 & 0.051 & 0.051 \\
      & 30,30 & 0.059 & 0.067 & 0.057 & 0.057 &     & 30,30 & 0.053 & 0.058 & 0.052 & 0.052 \\
  50  & 5,10  & 0.059 & 0.090 & 0.054 & 0.054 & 500 & 5,10  & 0.056 & 0.083 & 0.045 & 0.051 \\
      & 10,10 & 0.058 & 0.075 & 0.048 & 0.048 &     & 10,10 & 0.055 & 0.072 & 0.052 & 0.047 \\
      & 20,30 & 0.057 & 0.065 & 0.051 & 0.051 &     & 20,30 & 0.052 & 0.058 & 0.052 & 0.052 \\
      & 30,30 & 0.057 & 0.060 & 0.051 & 0.051 &     & 30,30 & 0.052 & 0.058 & 0.049 & 0.050 \\
  100 & 5,10  & 0.057 & 0.089 & 0.049 & 0.049 & 750 & 5,10  & 0.056 & 0.087 & 0.051 & 0.051 \\
      & 10,10 & 0.056 & 0.068 & 0.047 & 0.047 &     & 10,10 & 0.055 & 0.066 & 0.049 & 0.045 \\
      & 20,30 & 0.056 & 0.064 & 0.051 & 0.051 &     & 20,30 & 0.054 & 0.059 & 0.052 & 0.049 \\
      & 30,30 & 0.055 & 0.064 & 0.052 & 0.052 &     & 30,30 & 0.050 & 0.055 & 0.053 & 0.053 \\
  \hline
\end{tabular}\caption{Estimated type I error for the Setting 1 ($d=5$)}  \label{tab:level:rev1}
\end{table}

\begin{table}
\centering

\begin{tabular}{cccccc|cccccc}
  $k$ & $n_{1r},n_{2r}$ &  Test 4 & Test 5 & Test 6 & Test 7 & $k$ & $n_{1r},n_{2r}$ & Test 4 & Test 5 & Test 6 & Test 7 \\ \hline
  20  & 5,10  & 0.032 & 0.091 & 0.030 & 0.031 & 200 & 5,10  & 0.027 & 0.085 & 0.025 & 0.025 \\
      & 10,10 & 0.045 & 0.081 & 0.037 & 0.035 &     & 10,10 & 0.036 & 0.068 & 0.031 & 0.032 \\
      & 20,30 & 0.048 & 0.064 & 0.044 & 0.045 &     & 20,30 & 0.043 & 0.065 & 0.045 & 0.045 \\
      & 30,30 & 0.051 & 0.063 & 0.048 & 0.050 &     & 30,30 & 0.051 & 0.059 & 0.046 & 0.046 \\
  50  & 5,10  & 0.027 & 0.097 & 0.031 & 0.031 & 500 & 5,10  & 0.026 & 0.087 & 0.029 & 0.029 \\
      & 10,10 & 0.040 & 0.078 & 0.033 & 0.034 &     & 10,10 & 0.033 & 0.071 & 0.033 & 0.033 \\
      & 20,30 & 0.046 & 0.067 & 0.047 & 0.044 &     & 20,30 & 0.046 & 0.060 & 0.041 & 0.040 \\
      & 30,30 & 0.047 & 0.062 & 0.047 & 0.047 &     & 30,30 & 0.045 & 0.055 & 0.046 & 0.046 \\
  100 & 5,10  & 0.029 & 0.090& 0.032 & 0.032 & 750 & 5,10  & 0.027 & 0.087 & 0.027 & 0.027 \\
      & 10,10 & 0.040 & 0.075 & 0.035 & 0.035 &     & 10,10 & 0.037 & 0.071 & 0.032 & 0.032 \\
      & 20,30 & 0.046 & 0.064 & 0.042 & 0.042 &     & 20,30 & 0.042 & 0.060 & 0.041 & 0.041 \\
      & 30,30 & 0.047 & 0.062 & 0.044 & 0.044 &     & 30,30 & 0.043 & 0.053 & 0.048 & 0.049 \\
  \hline
\end{tabular}\caption{Estimated type I error for the Setting 1 ($d=10$). \label{tab:level:rev2}}
\end{table}

\begin{table}
\centering

\begin{tabular}{cccccc|cccccc}
  $k$ & $n_{1r},n_{2r}$ &  Test 4 & Test 5 & Test 6 & Test 7 & $k$ & $n_{1r},n_{2r}$ & Test 4 & Test 5 & Test 6 & Test 7 \\ \hline
  20  & 5,10  & 0.010 & 0.097 & 0.011 & 0.016 & 200 & 5,10  & 0.006 & 0.090 & 0.011 & 0.011 \\
      & 10,10 & 0.019 & 0.085 & 0.019 & 0.019 &     & 10,10 & 0.017 & 0.074 & 0.017 & 0.017 \\
      & 20,30 & 0.034 & 0.069 & 0.032 & 0.032 &     & 20,30 & 0.031 & 0.061 & 0.031 & 0.031 \\
      & 30,30 & 0.035 & 0.062 & 0.035 & 0.034 &     & 30,30 & 0.036 & 0.059 & 0.032 & 0.032 \\
  50  & 5,10  & 0.008 & 0.094 & 0.014 & 0.013 & 500 & 5,10  & 0.007 & 0.084 & 0.012 & 0.012 \\
      & 10,10 & 0.019 & 0.077 & 0.016 & 0.017 &     & 10,10 & 0.018 & 0.070 & 0.016 & 0.016 \\
      & 20,30 & 0.034 & 0.062 & 0.031 & 0.031 &     & 20,30 & 0.031 & 0.058 & 0.030 & 0.030 \\
      & 30,30 & 0.036 & 0.058 & 0.035 & 0.035 &     & 30,30 & 0.033 & 0.058 & 0.034 & 0.034 \\
  100 & 5,10  & 0.007 & 0.089 & 0.009 & 0.010 & 750 & 5,10  & 0.008 & 0.082 & 0.013 & 0.013 \\
      & 10,10 & 0.019 & 0.072 & 0.016 & 0.016 &     & 10,10 & 0.016 & 0.070 & 0.016 & 0.016 \\
      & 20,30 & 0.030 & 0.061 & 0.031 & 0.031 &     & 20,30 & 0.029 & 0.062 & 0.030 & 0.030 \\
      & 30,30 & 0.040 & 0.058 & 0.033 & 0.033 &     & 30,30 & 0.038 & 0.059 & 0.036 & 0.036 \\
  \hline
\end{tabular}\caption{Estimated type I error for the Setting 1 ($d=20$).  \label{tab:level:rev3}}
\end{table}

\begin{table}
\centering

\begin{tabular}{ccccccccc}
  $d$ & $k$   & Test 1 & Test 2 & Test 3 & Test 4 & Test 5 & Test 6 & Test 7  \\ \hline
  5   & 50  & 0.003 & 0.008 & 0.010 & 0.002 & 0.002 & 0.005 & 0.075  \\
      & 200 & 0.024 & 0.025 & 0.033 & 0.007 & 0.008 & 0.014 & 0.302 \\
      & 750 & 0.089 & 0.088 & 0.088 & 0.042 & 0.043 & 0.045 & 1.015\\
  10  & 50  & 0.006 & 0.009 & 0.003 & 0.003 & 0.003 & 0.003 & 0.078  \\
      & 200 & 0.029 & 0.025 & 0.028 & 0.010 & 0.012 & 0.014 & 0.305 \\
      & 750 & 0.092 & 0.090 & 0.088 & 0.032 & 0.037 & 0.045 & 1.132\\
  20  & 50  & 0.014 & 0.006 & 0.010 & 0.004 & 0.005 & 0.008 & 0.106 \\
      & 200 & 0.036 & 0.030 & 0.032 & 0.016 & 0.016 & 0.017 & 0.404 \\
      & 750 & 0.127 & 0.114 & 0.115 & 0.057 & 0.059 & 0.064 & 1.514\\
  \hline
\end{tabular}\caption{Average of the CPU time to calculate the test statistics  for $n_{1r}=n_{2r}=30$.} \label{tab:cpu}
\end{table}

\end{document}